\documentclass[final,10pt]{elsart}
\pdfoutput=1
\usepackage{amssymb,latexsym,amsmath,amsfonts}
  \usepackage[dvips]{graphicx}
\usepackage{subfigure}
\usepackage{hyperref}

\usepackage{pdfpages}

\newcommand{\pardt}{\partial t}
\newcommand{\pardx}{\partial x}

\newcommand{\pardy}{\partial y}

\newcommand{\be}{\begin{equation}}
\newcommand{\ee}{\end{equation}}
\renewcommand {\theequation}{\thesection.\arabic{equation}}

\begin{document}

\begin{frontmatter}

\title{High Reynolds number Navier-Stokes solutions and
boundary layer separation induced by a rectilinear vortex}
\author[autore]{F. Gargano}
\ead{gargano@math.unipa.it}
\author[autore]{M.Sammartino}
\ead{marco@math.unipa.it}
\author[autore]{V.Sciacca}
\ead{sciacca@math.unipa.it}
\maketitle
\date{}

\address[autore]{Dept of Mathematics,  University of Palermo\\ Via Archirafi 34, 90123
Palermo, Italy.}

\begin{abstract}
We compute the solutions of Prandtl's and Navier-Stokes equations for
the two dimensional flow induced by a rectilinear vortex interacting with
a boundary in the half plane.
For this initial datum Prandtl's equation develops, in a finite  time,
a separation singularity.
We investigate the different stages of unsteady separation for Navier-Stokes solution
at different Reynolds numbers $Re=10^3-10^5$, and we show the presence of a
large-scale interaction between the viscous boundary layer and the inviscid outer flow.
We also see a subsequent stage, characterized by the presence of a small-scale
interaction, which is visible only for moderate-high Re numbers $Re=10^4-10^5$.
We also investigate the asymptotic validity of boundary layer theory by
comparing Prandtl's solution to Navier-Stokes solutions during the various
stages of unsteady separation.
\end{abstract}


\end{frontmatter}

%%%%%%%%%%%%%%%%%%%%%%%%%%%%%%%%%%%%%%%%%%%%%%%%%%%%%%%%%%%%%%%%%%%%%%%%%%%%%%%%%%%%%
\section{Introduction}
\setcounter{equation}{0}

The aim of this paper is to analyze  the unsteady separation process
of a 2D incompressible Navier-Stokes (NS) flow induced by the interaction of a point
vortex with a physical boundary.
We shall solve Navier-Stokes equations at different $Re$ regimes
($Re=10^3-10^5$) and we shall compare these results with the predictions of
the classical Boundary Layer Theory (BLT) as expressed by Prandtl's equations.

\subsection{Prandtl's equation}
Prandtl's equations can be derived from the Navier-Stokes equations as the formal
asymptotic limit for $Re\rightarrow \infty$.
It is well known how, for many significant flows, Prandtl's solutions develop
a singularity  (see for example  \cite{vDS80,Cas00,PSW91a,PSW91b} and
\cite{vDC90,DW84}).

In their seminal work devoted to the analysis of the flow around the impulsively
started disk,  Van Dommelen and Shen \cite{vDS80}  found that, in a finite time,
a singularity in the Prandtl solution forms (the VDS singularity).
The difficulties that had prevented  the previous investigations to give reliable
results on the singularity,  were solved in \cite{vDS80} using a Lagrangian formulation,
which allowed to overcome the problem of the growth in time of the normal velocity
component $V$ i.e. the growth in time of the boundary layer.
More recently the same problem has been tackled in \cite{DLSS06,GSS09,GLSS09} where,
using  a high resolution Eulerian spectral method,  the authors have tracked the VDS
singularity in the complex plane before the real blow up of the solution, and have
classified it as a cubic-root singularity.

According to the Moore-Root-Sears (MRS) model the singularity in the solution of  Prandtl's
equation is related to the unsteady separation of the boundary layer, see also
\cite{ST75}.
In fact the occurrence of a singularity means that the normal component of the
velocity  $V$ becomes infinite  with the relative ejection of vorticity and flow
particles from within the boundary layer into the outer flow, with the consequent
breakdown of the assumptions which Prandtl's equation are based on.

Before the occurrence of the singularity, the adverse streamwise
pressure gradient imposed across the boundary layer induces the
formation of a back-flow region.
It has been observed that, generally, the formation of a recirculation region
corresponds to the vanishing of the vorticity at a point of the boundary.
The back flow region grows in time in the
streamwise direction, and ejects farther in the normal direction.
This results in the formation of a sharp spike in the displacement thickness and in
the streamlines. The spike in the streamlines, at the singularity time, reaches the
outer external flow, once again a signal of the interaction of the boundary layer with
the outer flow.

As an historical remark we mention that before the important results obtained by
Van Dommelen $\&$ Shen, the classical definition of unsteady separation was connected
with the formation of reversed flow and the vanishing of the wall shear.
However Sears and Telionis, in \cite{ST75}, observed that the presence of reversed flow
is not in itself sufficient to lead to unsteady separation; they quoted examples
of flows with vanishing wall shear for which a breakaway is never expected to occur.

An interesting  review on boundary layer theory and on the many numerical experiments
which followed Van Dommelen and Shen's work is given by Cowley in \cite{Cow01}.
For the reader interested in the results of the mathematical theory of the Prandtl
equations, see \cite{CS00}.

\subsection{Navier-Stokes solutions and comparison with BLT}

In the rest of the paper we shall denote the Reynolds number with
$Re$ defined as:
$$
Re=aU_{c}/\nu
$$
where $a$ and $U_c$ are the distance from the wall and the velocity with respect to
the wall of the point vortex, while $\nu$ is the kinematic viscosity.
Solving the Navier-Stokes equations at different Reynolds numbers
we shall follow the unsteady separation process.
We shall find significant differences in
the behavior of NS solutions between low ($10^3\leq Re \leq 2 \cdot 10^3$) and moderate-high
($3�\cdot 10^3 \leq Re \leq 10^5$) Reynolds number regimes. In fact we shall see how, at
different $Re$, different kinds of interactions will establish between the viscous
boundary layer and the outer flow; we shall also see that these interactions are
the responsible for the ultimate failure of the Prandtl's equations to give an
accurate approximation of the NS flow at the $Re$ numbers we have tested.

In the classical BLT the streamwise pressure gradient is imposed by the outer flow while
the normal pressure gradient is zero to leading order.
Therefore through the analysis of topological changes of the pressure gradient,
we shall be able to distinguish the different  stages of the interaction between the
BL and  the outer flow.
Moreover the evolution of the pressure gradient will give indications on the agreement
between the Prandtl and the NS solutions.

During the early stage we shall observe that the Prandtl solution is quite close,
both qualitatively and quantitatively, to the  Navier-Stokes solutions.

On the other hand relevant discrepancies can be observed  when the boundary layer flow
starts interacting with the outer flow over a large streamwise scale.
This event can be related to the formation of an inflection point in the streamwise
pressure gradient.
This phenomenon is observed for all $Re$ we have considered.

A second different interaction occurring on a smaller scale  is present only for
moderate-high $Re$ numbers.
We notice that, differently from what happens for lower $Re$ numbers,
several local maxima-minima form in the streamwise pressure gradient,
forcing the formation of several recirculation regions and of strong
gradients in the solution.
The first appearance of spiky-behavior in the streamlines and vorticity contour level
signals the beginning of this new stage.
At this time any comparison with Prandtl's solution fails, even if the formation of
large gradients in the solution resembles the terminal singularity stage of Prandtl's
solution.
Both types of interaction (large scale and small scale) begin quite early with respect
to the first viscous-inviscid interaction that occurs in Prandtl's solution.

In the literature there are several attempts  to incorporate the interaction between the
boundary layer and the outer flow in a theory that would improve the classical BLT.
We mention the work in \cite{PSW91b} where the authors assume that, as the spike in the
displacement thickness grows, the outer flow begins to respond to the boundary layer.
However the solutions of these Interactive Boundary Layer Theory terminate with
a singularity at a time prior to Prandtl's singularity time.
See also \cite{CC89} and the discussion in \cite{CSW96}.
A possible cure to this was proposed in \cite{HSW91} and \cite{LWBS98} where the
effects of an increasing normal pressure gradient (which is considered constant in
the classical BLT) are taken into account.
However none of the theories trying to go beyond the classical BLT is fully satisfactory
and the problem of a coherent asymptotic theory able to describe the BL separation
phenomena is still an open problem, see the discussion in \cite{OC02} and the
review paper \cite{Cow01}.

\subsection{Plan of the paper}

In the next Section we introduce the physical problem, a 2D vortex
interacting with a wall, and discuss the regularization procedure we have
adopted to treat the relative initial datum.
In Section 3 we present the numerical schemes we have used to solve Prandtl's  and
Navier-Stokes equations.
The numerical results obtained from Prandtl's equation up to singularity formation are
briefly (as this problem was already discussed in great detail in \cite{Wal78,PSW91a})
described in Section 4.
In  Sections 5 and 6, we show the results obtained for the Navier-Stokes solutions,
and we analyze the different stages of unsteady separation.
In particular the large-scale interaction stage, that develops for all $Re$ numbers we
considered, is discussed in Section 5, while the small-scale
interaction stage, found for moderate-high
is discussed in Section 6.
Our analysis follows the treatment of \cite{Cas00,OC02}
where the authors studied the interaction of the thick core vortex with a
boundary and confirms the scenario described in these paper, as well in \cite{KCH07}.

In Section 7 we shall discuss in more detail the physical phenomena leading to
the different kind of interactions described in the previous Sections;
in particular we shall see first the formation of dipolar vortical structures as
the signal of the small scale interaction and second a significant increase in
the enstrophy of the flow as the result of the movement of these dipolar structures
toward the wall.
This analysis is influenced by the findings appeared in a recent series of papers
\cite{CH02,CB06,KCH07} and previously in \cite{OR90,CL91},
where the case of the collision of a dipole vortex with a boundary was considered.

%%%%%%%%%%%%%%%%%%%%%%END INTRODUCTION%%%%%%%%%%%%%%%%%%%%%%%%%%%%%%%%%%%%%%%

\section{Statement of the problem}
\setcounter{equation}{0}

The initial fluid configuration consists of a point-vortex immersed in a 2D viscous
incompressible flow  at rest at infinity and  bounded by an infinite rectilinear wall.
The vortex is placed at a distance $a$ from the wall, and is taken
with positive rotation and strength $k$.
In the inviscid case, the vortex moves to the right parallel to the wall with
constant velocity $U_{c}=k/4a\pi$.
We refer to \cite{LAMB} for more details.

We introduce a cartesian frame $(x,y)$, such that the $x-$axis coincides with the
solid boundary.
The point-vortex is centered in $(0,a)$ and we denote by $u$ and $v$  the streamwise
and normal velocity components respectively.
As in \cite{Wal78,PSW91a}, by superimposing a uniform tangential velocity
equal and opposite to $U_c$, we shall study the system in the frame comoving with the
vortex, so that the wall moves with constant  velocity $-U_c$.
We introduce dimensionless variables  taking the distance of the vortex from the
wall $a$ and the velocity $U_c$ as, respectively, characteristic length and velocity.

The governing equations for the flow evolution are the Navier-Stokes equations
in the domain $(-\infty,\infty)\times[0,\infty)$ that write as:
\begin{eqnarray}
\frac{\partial u}{\pardt}+u\frac{\partial u}{\pardx}+v\frac{\partial u}{\pardy}+
\frac{\partial p}{\pardx}&=&
\frac{1}{Re}\Delta u, \label{NSequation_u}\\
\frac{\partial v}{\pardt}+u\frac{\partial v}{\pardx}+v\frac{\partial v}{\pardy}+
\frac{\partial p}{\pardy}&=&
\frac{1}{Re}\Delta v, \label{NSequation_v}\\
\frac{\partial u}{\pardx}+\frac{\partial v}{\pardy}&=& 0 \label{NSequation_incomp} \; .
\end{eqnarray}
The boundary conditions are:
\begin{eqnarray}
& & u=-1 \quad v=0, \qquad \mbox{at} \quad y=0\, , \\
& & u=-1 \quad v=0, \qquad \mbox{when} \quad x\rightarrow\pm\infty \, ,  \\
& & u=-1 \quad v=0, \qquad \mbox{when} \quad y\rightarrow\infty\; .
\end{eqnarray}
The initial data for the velocity components are:
\begin{equation}
u_0=\partial_y \psi_E \quad {\rm and} \quad v_0=-\partial_x
\psi_E,\label{init_Prandtl}
\end{equation}
where
\begin{equation}
\psi_E(x,y)=-\log\left(\frac{x^2+(y-1)^2}{x^2+(y+1)^2}\right)-y \label{psi}
\end{equation}
is the streamfunction of the inviscid steady Euler solution for this configuration.

The no-slip boundary condition imposed at the wall leads to vorticity generation at
the boundary which develops the unsteady separation phenomenon.
To describe the flow inside the boundary-layer, one defines the scaled
normal coordinate $Y$ and normal velocity  $V$  by the well known boundary layer
scaling: $y=\sqrt{Re}Y$ and $v=\sqrt{Re}V$.
Prandtl's equations are obtained, to first order, introducing the above
scaling into the Navier-Stokes equations.
For the rectilinear vortex case, Prandtl's equations are:
\begin{eqnarray}
\frac{\partial u}{\partial t}+u\frac{\partial u}{\partial x}+V
\frac{\partial u}{\partial Y}-U_{\infty}\frac{\partial
U_{\infty}}{\partial x}&=&\frac{\partial^2 u}{\partial Y^2}, \label{pramomentum}\\
\frac{\partial u}{\partial x}+\frac{\partial V}{\partial Y}&=&0,\label{consmass}
\end{eqnarray}
with initial datum and boundary conditions given by
\begin{eqnarray}
&&u(x,Y,0)=U_{\infty},\label{initialprandtl}\\
&& u(x,0,t)=-1, \quad u(x,Y\rightarrow\infty,t)=U_{\infty},\label{boundaryprandtl}
\end{eqnarray}
where $U_\infty=-1+4/(x^2+1)$ is the inviscid solution at the boundary on
the reference frame comoving with the vortex.

%%%%%%%%%%%%%%%%%%%%%%%%%%%%%%%%%%%%%%%%%%%%%%%%%%%%%%%%%%%%%%%%%%%%%%%%%%%%%%%%%%
\section{The numerical schemes}
\setcounter{equation}{0}

\subsection{Numerical schemes for Prandtl's equations}
In this section we explain the numerical method used to solve the boundary-layer
equations \eqref{pramomentum}-\eqref{consmass} with initial and boundary conditions
\eqref{initialprandtl}-\eqref{boundaryprandtl}.
This problem was first investigated by Walker in \cite{Wal78},
and later by Peridier, Smith and Walker in \cite{PSW91a}.
The problematic numerical instabilities developed by the numerical method of
\cite{Wal78} in Eulerian formulation, were overcome using in \cite{PSW91a}  a
Lagrangian formulation, and using an ADI scheme with upwind-downwind differencing
approximation for the convective terms.
Probably the use of a central differencing approximation for first derivatives and
the lack of the necessary resolution were the main reasons causing
instabilities in \cite{Wal78}.
Our simulation are based on a Runge-Kutta IMEX scheme for time advancing and on a
two step Richtmyer-Lax-Wendroff approximation for the convective terms,
which makes possible to carry out the computation almost up to the singularity time
in the conventional Eulerian formulation.

Following \cite{Cas00} we shall map the physical domain  $(-\infty,\infty)\times[0,\infty)$
onto the finite domain  $(-1,1)\times[0,1)$.
The map is explicitly given by the following transformations:
\begin{equation}
\hat{x}=\frac{2}{\pi}\arctan{\left(\frac{x-x_{s}}{\alpha}\right)}, \qquad
\hat{y}=\frac{2}{\pi}\arctan{\left(\frac{Y}{\beta}\right)}, \label{transf}
\end{equation}
being $x_{s}$ the streamwise location where the singularity forms,
and $\alpha$ and $\beta$ are positive parameters.
These transformations cluster the computational grid close $(x_{s},0)$;
the parameters $\alpha$ and $\beta$ determine  the degree of focusing of the grid.
We note however that as the normal velocity component $V\rightarrow\infty$
as $Y\rightarrow\infty$, we need to truncate the normal domain to a value
$Y_M$.
The value $Y_M$ (where the boundary condition  $u(x,Y_M,t)=U_\infty$ is
imposed) must be big enough so that the growth of the boundary layer does not
affect the solution at $Y_M$.
In our calculation we find that the value $Y_M=20$  is enough to ensure the
reliability of the computed solution up to times very close to the singularity.
Therefore the computational normal domain is $[0,\hat{y}(Y_M)]$.

Using all these transformation Prandtl's equations
\eqref{pramomentum}--\eqref{boundaryprandtl} become:
\begin{eqnarray}
\frac{\partial u}{\pardt}+u\frac{\partial \hat{x}}{\partial x} \frac{\partial u}{\partial \hat{x}}+
V\frac{\partial \hat{y}}{\partial Y}\frac{\partial u}{\partial \hat{y}}
-U_{\infty}\frac{\partial U_{\infty}}{\partial x}&=&
\left(\frac{\partial \hat{y}}{\partial Y}\right)^2\frac{\partial^2 u}{\partial \hat{y}^2}+\frac{\partial^2 \hat{y}}{\partial Y^2}\frac{\partial u}{\partial \hat{y}}, \label{praeq}\nonumber \\ \\
\frac{\partial \hat{x}}{\partial x}\frac{\partial u}{\partial \hat{x}}+
\frac{\partial \hat{y}}{\partial Y}\frac{\partial V}{\partial \hat{y}}&=&0,\label{incompt}\\
u(\hat{x},\hat{y},0)&=&U_{\infty}   ,\label{initialprat}\\ %&=&-1+\frac{4}{x^2+1}
u(\hat{x},0,t)=-1, \quad u(\hat{x},\hat{y}(Y_M),t)&=&U_{\infty}, \label{boundaryprat} \\
u(\hat{x}\rightarrow\pm 1,\hat{y},t)&=&-1.
\end{eqnarray}
The normal velocity $V$ is computed  from the incompressibility equation \eqref{incompt}
through numerical integration.
In Eq.\eqref{praeq} the  convective term
is approximated by the two step Richtmyer-Lax-Wendroff
rule (see \cite{Lev92} for details).
The diffusive term is approximated by the usual 3-point rule.
For the temporal discretization we use the Runge-Kutta IMEX
midpoint method (2,3,3) (see \cite{ARS97} for details), and the
relative Runge-Kutta tables are:

\be
\begin{tabular}{l|lll}
0        & 0            & 0           & 0 \\
$\gamma$   & $\gamma$   & 0           & 0 \\
$1-\gamma$ & $\gamma-1$ &$2(1-\gamma)$& 0 \\
\hline
         & 0            &$1/2$        &$1/2$
\end{tabular}\qquad \quad
\begin{tabular}{l|ll}
$\gamma$   & $\gamma$   & 0      \\
$1-\gamma$ &$1-2\gamma$ & $\gamma$ \\
\hline
         &    1/2   & 1/2
\end{tabular}
\label{233}
\ee
with $\gamma=(3+\sqrt{3})/6$.
We have used a grid of $8192\times2400$ mesh points clustered in ($0.21,0$),
and we have set the parameters $\alpha=0.1,\beta=0.1$ in the domain-transformation
function \eqref{transf}.
The time step $\Delta t$ changes at each step according to the
CFL condition $\Delta t=\min(\Delta \hat{x}/|u {\partial_x \hat{x}}|,
\Delta \hat{y}/|V {\partial_Y \hat{y}}|)$,
where $\Delta \hat{x}$ and $\Delta \hat{y}$ are the tangential and normal mesh sizes.
This preserves the computation from any numerical instability until the formation of
the singularity.

A different issue (which is present in most of the Prandtl computations presented
in the literature) is the incompatibility between the boundary condition at $Y=0$
\eqref{boundaryprandtl} and the initial condition \eqref{initialprandtl}.
The diffusion adjusts this incompatibility in zero time, and in fact
our finite-difference code  smooths out the discontinuity
during the first time step, and no instability occurs.
One has to check that the numerical results for the flow evolution do not depend
on the way the code smooths out the discontinuity. The procedure we have adopted
is the following. We have initiated the flow imposing an initial datum
which interpolates smoothly and monotonously between the value of the velocity at the boundary
and $U_\infty$. The interpolation occurs in a small layer of size $\epsilon$.
We have seen that (after a transient) the flow evolution is independent from
$\epsilon$, the only difference being a time shift $\Delta T_\epsilon$
with respect to the solution obtained imposing the discontinuous initial
datum.
We have seen that $\Delta T_\epsilon\sim\epsilon^2$ which is consistent
with the fact that the imposition of the artificial interpolation layer
has the same effect of the initial diffusive layer.
This behavior is independent of the chosen interpolating functions.
The same procedure was adopted for the Navier-Stokes solutions.

\subsection{Numerical schemes for Navier-Stokes equations}

We solve the equations \eqref{NSequation_u}--\eqref{NSequation_incomp}
in the vorticity-streamfunction formulation:
\begin{eqnarray}
\frac{\partial \omega}{\pardt}+u\frac{\partial \omega}{\pardx}+v\frac{\partial \omega}{\pardy}&=&
\frac{1}{Re}\Delta \omega, \label{NSequation}\\
\Delta \psi&=&-\omega, \label{poisson}\\
u=\frac{\partial \psi}{\pardy}, \quad v&=&-\frac{\partial \psi}{\pardx},\label{velocityew}\\
\omega(x,y,t=0)=\omega_0&=&4\pi\delta_{(0,1)}, \label{NSinit}\\
\omega(x\rightarrow\pm\infty,y,0)=\omega(x,y\rightarrow\infty,0)&=&0, \label{vorticitybc}\\
u(x,0,t)=-1,\quad v(x,0,t)&=&0 . \label{noslipppp}
\end{eqnarray}
As the initial datum is singular, we convolve with the mollifier
$\phi_{\sigma}(x,y)=\frac{1}{\sigma^{2}}e^{-(x^2+y^2)/\sigma^2}$,
obtaining the regularized initial datum
$$
{\omega_{0}}_{\sigma}=\omega_0\ast\phi_{\sigma}=\frac{4\pi}{ \sigma^{2}}e^{-(x^2+(y-1)^2)/\sigma^2}\,.
$$
We have chosen the value $\sigma=0.05$ and our results do not
depend on this particular choice.
We have checked that in the boundary  layer and up to the computational
time $T=1.5$,  the differences  of the velocity fields
between the cases $\sigma=0.05$ and $\sigma=0.025$, are below the precision of the machine.

We have used a stretching function clustering the grid
both near the solid boundary and at the point $(0,1)$  where the vortex blob is located.
The stretching functions are:
\begin{eqnarray}
\bar{x}&=&\alpha_x \arctan\left(\frac{x-x_1}{\beta_{x_1}}\right)\left[\gamma_x+
\arctan\left(\frac{x-x_2}{\beta_{x_2}}\right)\right], \quad x_1<  x_2 \label{mapx}\\
\bar{y}&=&\alpha_y \arctan\left(\frac{y-y_2}{\beta_{y_2}}\right)\left[\gamma_y+
\arctan\left(\frac{y-y_1}{\beta_{y_1}}\right)\right], \quad y_1 > y_2 \label{mapy}
\label{clustNS}
\end{eqnarray}
which map the domain $(-\infty,+\infty)\times[0,\infty)$ in
$\left(-\alpha_{x}\frac{\pi}{2}(\gamma_x-\frac{\pi}{2}),
\alpha_{x}\frac{\pi}{2}(\gamma_x+\frac{\pi}{2})\right)
\times\left[0,\alpha_{y}\frac{\pi}{2}(\gamma_y+\frac{\pi}{2})\right)$.
In \eqref{mapx} (with analogous meaning for \eqref{mapy}) $\beta_{x_1}$ and
$\beta_{x_2}$ are positive parameters tuning the strength of the focusing of
the grid points close to $x_1$ and $x_2$ respectively,
$\alpha_x$ is a normalizing factor, and $\gamma_{x}$ is a parameter
chosen to make the function \eqref{mapx} bijective.
In particular to have the bijective condition satisfied $\gamma_x$ must
be chosen so that:

\begin{equation}
\gamma_{x}>\frac{-\beta_{x_2}\beta_{x_1}^2-
\beta_{x_2}(x-x_{1})^2}{\beta_{x_1}\beta_{x_2}^2+\beta_{x_1}(x-x_{2})^2}
\arctan(\frac{x-x_{1}}{\beta_{x_1}})-\arctan\left(\frac{x-x_2}{\beta_{x_2}}\right)
\qquad \forall x \, . \label{bijective}
\end{equation}
The value  $\gamma_x=\gamma_y=5$ and $x_1=0,y_1=1,y_2=0$ in \eqref{mapx}--\eqref{mapy}
are kept fixed in all simulations, while the other parameters  are listed
in Table 1, where it is also reported the typical time-step size we had to adopt
to ensure stability of our procedure.\\

\begin{table}[ht]
\label{comp_param}
\caption{Computational parameters.}
\begin{center}
\hskip1cm \begin{tabular}{|c|c|c|c|c|c|c|c|c||c|c|c|}\hline
$Re$   & Grid (x,y) &$x_2$ & $\beta_{x_1}$ & $\beta_{x_2}$  & $\beta_{y_1}$ & $\beta_{y_2}$  & dt\\
\hline
$10^3$ & 1025$\times$1025 & 0.4 & 0.6 & 0.15 & 0.2 & 0.5 & $7\cdot10^{-5}$\\
\hline
$10^4$ & 1025$\times$1025 & 0.33 & 0.6 & 0.1 & 0.05 & 0.5 & $2\cdot10^{-5}$\\
\hline
$10^5$ & 4097$\times$1025 & 0.24 & 0.65 & 0.085 & 0.02 & 0.4 & $2\cdot10^{-6}$\\
\hline
\end{tabular}
\end{center}
\end{table}
\bigskip

Applying \eqref{mapx}-\eqref{mapy} to Navier-Stokes equations
\eqref{NSequation}-\eqref{NSinit}, with the normalizing factors
$\alpha_{x}=(\frac{\pi}{2}(\gamma_x+\frac{\pi}{2}))^{-1}$ and
$\alpha_{y}=(\frac{\pi}{2}(\gamma_y+\frac{\pi}{2}))^{-1}$,
we obtain this system of equations to be solved in the finite domain
$(-\frac{\gamma_x-\frac{\pi}{2}}{\gamma_x+\frac{\pi}{2}},1)\times[0,1)$

\begin{eqnarray}
  \frac{\partial \omega}{\partial t} + \left(\frac{\partial{\bar{x}}}{\partial{x}}u-\frac{1}{Re}
\frac{\partial{^{2}\bar{x}}}{\partial{x^2}}\right)
\frac{\partial \omega}{\partial\bar{x}} +
  \left(\frac{\partial{\bar{y}}}{\partial{y}}v-
  \frac{1}{Re}\frac{\partial{^{2}\bar{y}}}{\partial{y^2}}\right)
  \frac{\partial \omega}{\partial \bar{y}}=
  \nonumber\\
  \frac{1}{Re}\left[\left(\frac{\partial{\bar{x}}}{\partial{x}}\right)^{2}
  \frac{\partial^2 \omega}{\partial \bar{x}^2} +
  \left(\frac{\partial{\bar{y}}}{\partial{y}}\right)^{2}
  \frac{\partial^2\omega}{\partial \bar{y}^2}\right]\, ,
  \label{vorticityeq} \\
  \left(\frac{\partial{\bar{x}}}{\partial{x}}\right)^{2}
  \frac{\partial^2 \psi}{\partial \bar{x}^2}+
  \left(\frac{\partial{\bar{y}}}{\partial{y}}\right)^{2}
  \frac{\partial^2 \psi}{\partial \bar{y}^2} +
  \frac{\partial{^{2}\bar{x}}}{\partial{x^2}}
  \frac{\partial   \psi}{\partial \bar{x}} +
  \frac{\partial{^{2}\bar{y}}}{\partial{y^2}}
  \frac{\partial\psi}{\partial \bar{y}} = -\omega \, ,
  \label{poissoneq} \\
\frac{\partial{\bar{y}}}{\partial y}\frac{\partial \psi}{\partial \bar{y}}=u, \qquad \frac{\partial{\bar{x}}}{\partial x}\frac{\partial
  \psi}{\partial x}=-v\, ,  \label{velocityeq} \\
\omega(x,y,0)=\frac{4\pi}{\sigma^{2}}e^{-(x^2+(y-1)^2)/\sigma^2} \, ,  \label{vorticityinit}\\
\omega(x\rightarrow\pm\infty,y,0)=\omega(x,y\rightarrow\infty,0)=0\, ,  \label{vorticityup}\\
u(x,0,t)=-1, \qquad v(x,0,t)=0 \, . \label{noslip}
\end{eqnarray}

Given that the streamfunction $\psi$ diverges when
$\bar{y}\rightarrow 1  (y\rightarrow\infty)$,
we truncate the physical domain to a value $Y_{max}$ where
the vorticity is  negligible (up to the
computational time).
We ran several experiments choosing different value $Y_{max}$,
and the value $Y_{max}=10$ was sufficient to avoid any possible dependency.
%All the results presented in this work are been tested to be grid independent.

Regarding advancing in time we have used a factored ADI
(Alternating-Direction-Implicit)
finite differences approximation, together with a  the Crank-Nicolson procedure,
which ensures a second order accuracy (see \cite{TAN}).
To find  the needed boundary condition for the vorticity at the boundary we have
adapted the well known Jensen's formula to non uniform grids \cite{EL96}.
Finally the Poisson's equation \eqref{poissoneq} is solved by a
$V-cycle$ multigrid iterative method with a
standard Gauss-Seidel colouring scheme smoother (see \cite{TAN,TRO}).

\section{Prandtl's solution}
\setcounter{equation}{0}

In this section we shall give a description of the physical phenomena occurring
in the boundary layer leading to the final break up of the solution due to the
blow-up of the first derivative of the streamwise velocity component.
In particular we shall describe the various stages leading to separation, and focus
our analysis on physical events like the formation of the recirculation
region and the first viscous-inviscid interaction.

These phenomena were already discussed in \cite{PSW91a, PSW91b} and therefore our
discussion will be brief and mainly focused on those elements useful for a
comparison with the Navier-Stokes solutions.
Our results agree with those presented in \cite{PSW91a, PSW91b} in the sense that
we find all the relevant phenomena related to the separation
process (the formation of the recirculation region, the formation of the zero
vorticity at the wall and the singularity) to occur at the same times
predicted in the mentioned papers.

\begin{figure}
\begin{center}
\includegraphics[width=12.cm,height=10.cm]{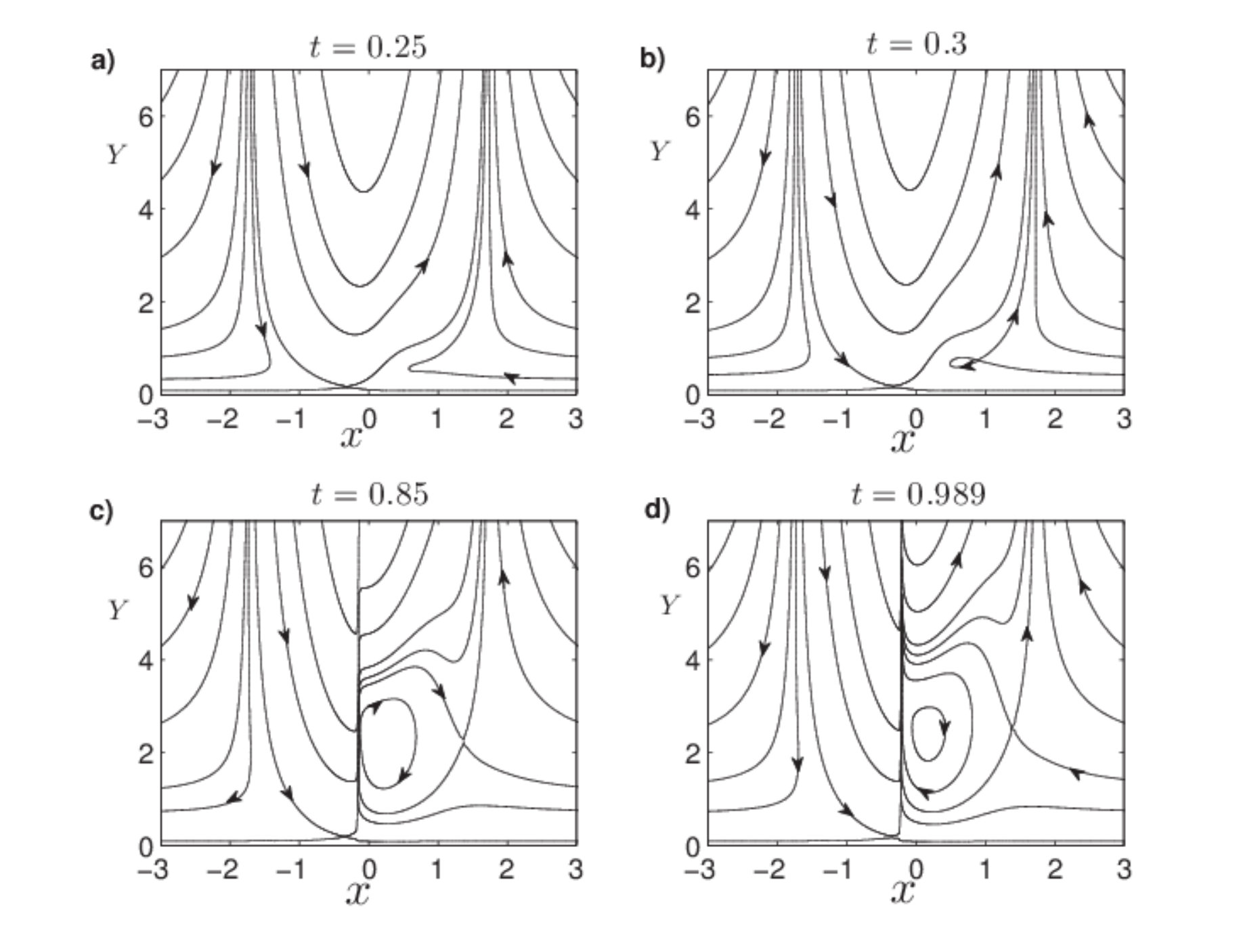}
\end{center}
   \caption{The streamlines of Prandtl's solution at different times.
   The recirculation region forms at $t\approx0.28$, and it is visible in b)
   at time $t=0.3$.
   At time $t=0.85$ a kink begins to form in the streamlines as the result of
   the first interaction of the boundary layer with the inviscid outer flow.
   At time $t=0.989$ a sharp spike forms in the streamwise location close to
   $x=-0.218$ as the result of the singularity.}
  \label{prastream}
\end{figure}

\begin{figure}
\begin{center}
 \includegraphics[width=11cm,height=6.cm]{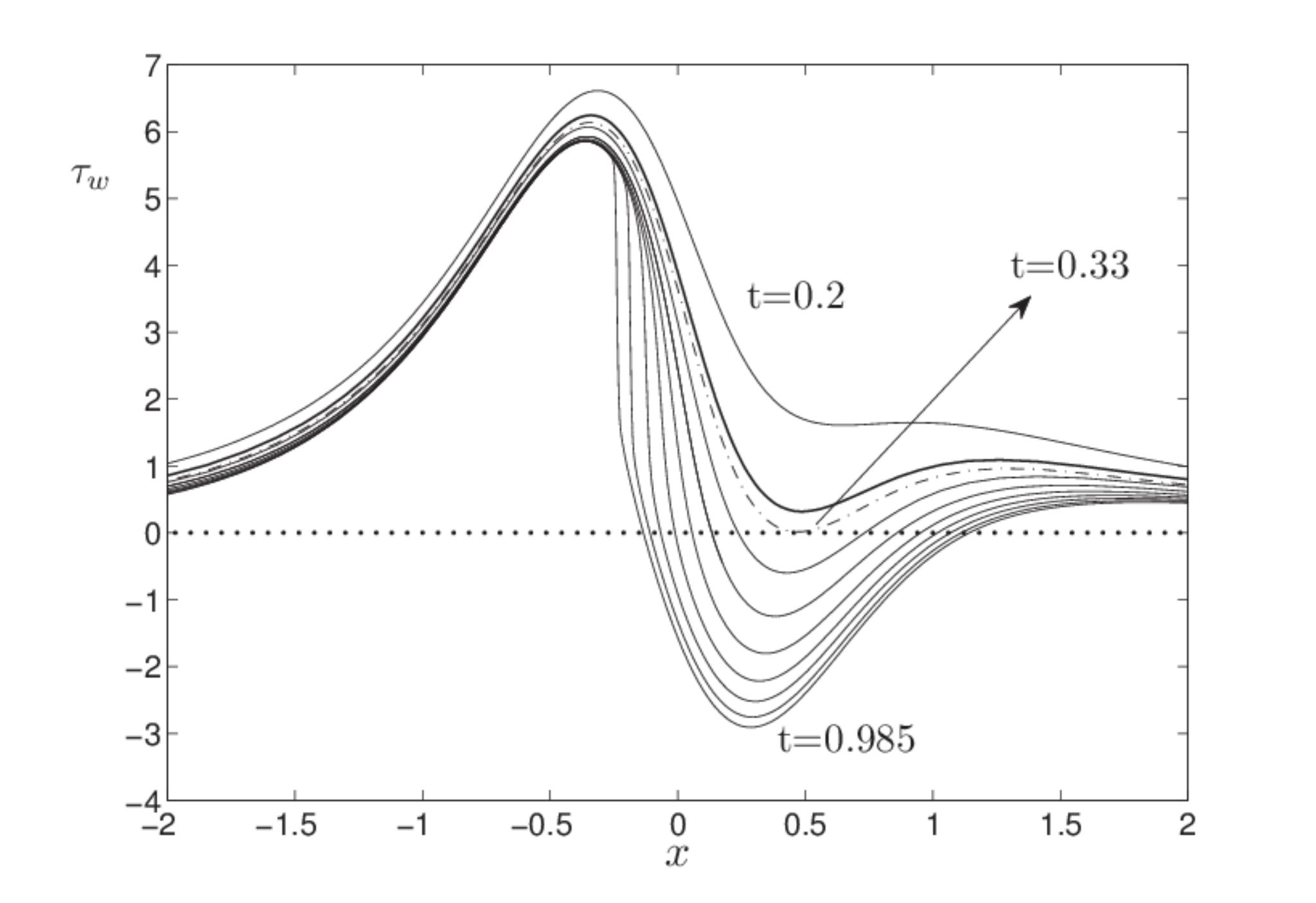}
 \end{center}
   \caption{The evolution of wall shear stress starting from time $t=0.2$
   up to $t=0.9$ (increments of 0.1) and 0.985.
   The dotted line is the wall shear at time $t=0.337$ when a region of negative
   values appears. }
  \label{wall_shear}
\end{figure}

At early stages the main phenomenon occurring in the boundary layer is the
generation of the vorticity at the boundary due to the no-slip boundary condition.
At the time $t_r\approx0.28$ the adverse pressure gradient, imposed by the outer flow,
leads to the formation of a recirculation region detached from the wall.
In fact, the formation of the recirculation region corresponds to the formation
of a stagnation point at $(x_r,y_r)=(0.74,0.659)$ as consequence of the vanishing
of the gradient of the streamfunction $\Psi$,
defined here by $\partial_Y \Psi=u$.
In Fig.\ref{prastream}b  the recirculation region
is clearly visible,  at time $t=0.3$, through the presence of a closed streamlines.
Notice that the time of creation of a recirculation region does not corresponds
to the time of  the vanishing of the wall shear
$\tau_w(x,t)=\frac{\partial u}{\partial Y}_{|Y=0}$,
which occurs at $t_w\approx0.337$ in $x_w\approx0.47$
(see Fig.\ref{wall_shear}), i.e. well after the formation of the recirculation region.
However this temporal gap disappears if one  considers the flow in the
laboratory frame where the recirculation region forms exactly at $t=t_w$.
The recirculation region, as time passes, grows and moves upstream.
At $t_k\approx0.85$ a kink forms in the streamlines, above and to the left of the recirculation region, see Fig.\ref{prastream}c,
due to the pressure gradient that forces the fluid to deflect upward.
According to the interpretation given in \cite{PSW91a,PSW91b},
the formation of the kink represents the first stage of the
viscous-inviscid interaction in the boundary layer.
In fact, for $t<t_k$ before the formation of the kink, the recirculation region,
although   significantly thickened in the streamwise  as well in the normal
direction, is still within the boundary layer, close to the wall.
For $t>t_k$, as the streamwise compression of the flow
pushes away fluid particles from the boundary,
the flow rapidly focuses in a narrow streamwise region to the left of the
recirculation region.
This fact reveals how the first stage of the viscous-inviscid interaction occurs.
The kink rapidly evolves in a sharp spike which is followed by the singularity
formation, due to the blow up of the first derivative of the streamwise
velocity component.
This final event occurs at time $t_s\approx0.989$ and at the spatial
location  $x_s\approx-0.218$.

\begin{figure}
\begin{center}
\includegraphics[width=11cm,height=6cm]{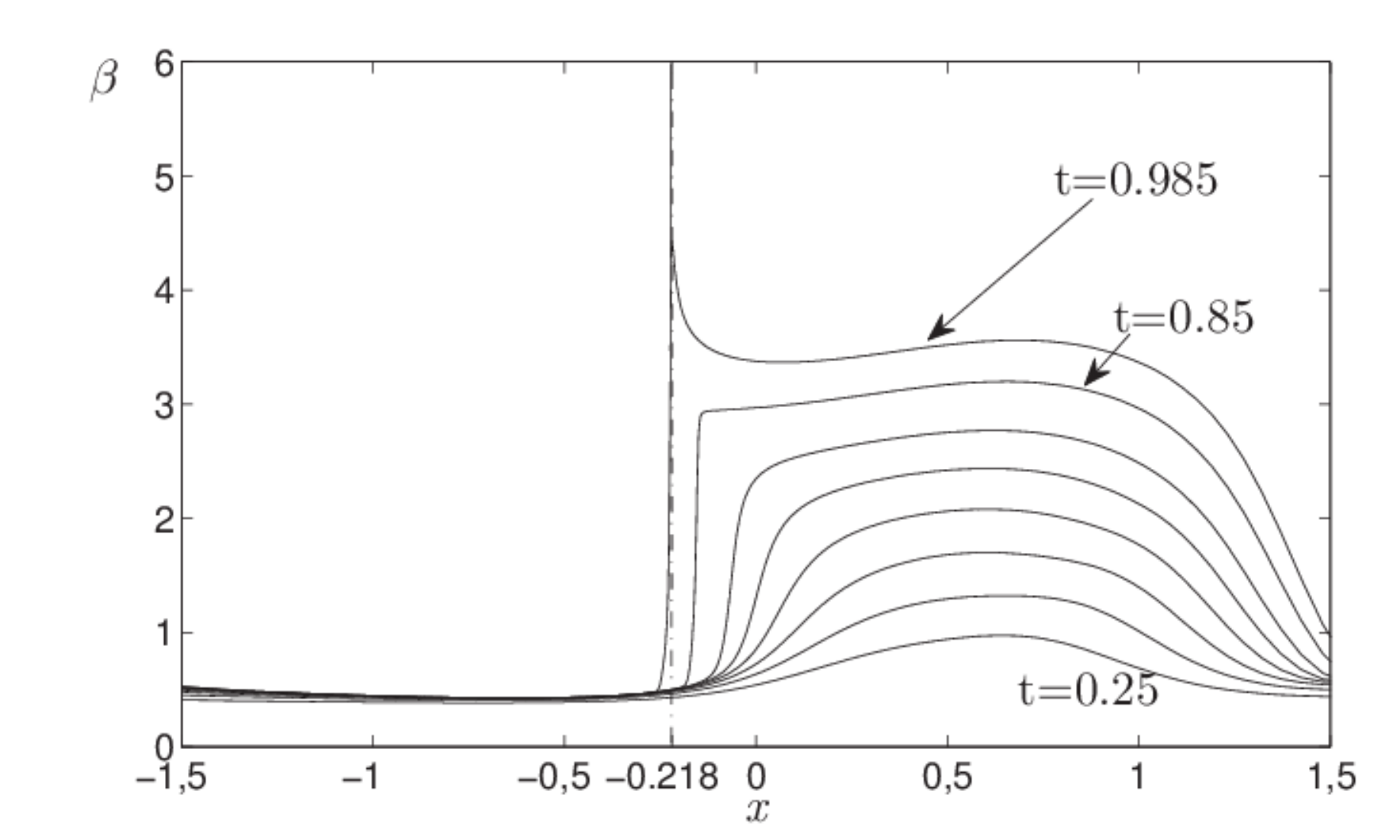}
\end{center}
\caption{The evolution in time of the displacement thickness from $t=0.25$
 up to $t=0.85$ (increments of 0.1) and at $t=0.985$. }
 \label{dispthick}
\end{figure}

The growth of the boundary-layer can be also illustrated through the displacement thickness, which is defined in  the laboratory  reference frame by:
\begin{eqnarray}
\beta(x)=\int_{0}^{\infty} \left(1-\frac{u+1}{U_{\infty}+1}\right)dY. \label{disp}
\end{eqnarray}
The time evolution of the displacement thickness is shown in Fig.\ref{dispthick}
from $t=0.25$ up to $t_k=0.85$ (with time-step of 0.1), and at time $t_s=0.985$.
Up to time $t_k$ the displacement thickness grows
in correspondence of the growth of the recirculation region.
At $t_k$ a local maximum (barely visible in the Figure) forms at $x\approx-0.14$,
and this signals the onset of interaction of the boundary layer flow with
the external flow.
Then the boundary layer abruptly focus in a narrow zone close to $x_s\approx-0.218$,
and at the singularity formation time $t_s=0.989$ its normal extension becomes
infinite in the boundary-layer scale,
leading to the final blow up of the displacement thickness.
Moreover the line of zero-vorticity focuses at the same streamwise
location as $\beta$ (not shown here).
This supports the MRS criterion for the boundary layer breakdown,
according to which singularity, and the consequent breakdown of
boundary layer assumption, occurs at the zero vorticity line.
Physically the singularity formation means that vorticity is ejected in the
outer flow from within the boundary layer.

\section{Navier-Stokes results: Large-Scale Interaction}
\setcounter{equation}{0}

In this section we shall study the behavior of the solutions of the Navier-Stokes equations at different $Re$ numbers ($10^{3}-10{^5}$).
We shall also be interested in the comparison between the Navier-Stokes solutions
and Prandtl's solution up to the singularity time $t_{s}=0.989$.
In particular we shall investigate the interaction  between the viscous
boundary layer and the inviscid outer flow occurring during the various stages of the unsteady boundary layer separation.

Before the beginning of this viscous-inviscid interaction the qualitative behavior of the
Navier-Stokes solution is similar for all the $Re$ numbers we have considered.
The main mechanisms driving the  behavior of the flow are the adverse pressure gradients
imposed by the vortex and the generation of vorticity at the boundary.
In this stage the first relevant phenomenon, visible for all the $Re$ numbers,
is the formation of a recirculation region, detached from the wall,
underneath and to the right of the central vortex.
The recirculation region forms at $t\approx0.28$, which is in perfect agreement
with Prandtl's solution.
The recirculation region starts to thicken  both in the streamwise and in the normal
direction, and the growth rate    depends on the $Re$ number:
the larger the $Re$ numbers the slower the recirculation region grows.
The wall shear stress $\tau_w$, defined here
as $\tau_w=-\omega_{|y=0}/\sqrt{Re}$,
vanishes at time close to $t=0.33$ for all the $Re$ numbers considered, well after the
formation of the recirculation region, as we have observed in Prandtl's solution.

The first relevant discrepancy between the  Navier-Stokes and  Prandtl's
solutions appears approximately when the  viscous-inviscid interaction  begins.
To analyze this interaction and  its influence on the flow evolution,
we consider the pressure gradient inside the boundary layer  as done in
\cite{OC02,Cas00} for the thick-core vortex case.
In fact for Prandtl's equation the streamwise pressure gradient
does not change as it is imposed by the outer flow, and the normal pressure
gradient is always zero.
Therefore we shall consider the variations of the pressure gradient of Navier-Stokes
equation as an indicator of the discrepancy between the Navier-Stokes
and Prandtl's solutions.

\begin{figure}
\begin{center}
\includegraphics[width=11cm,height=6cm]{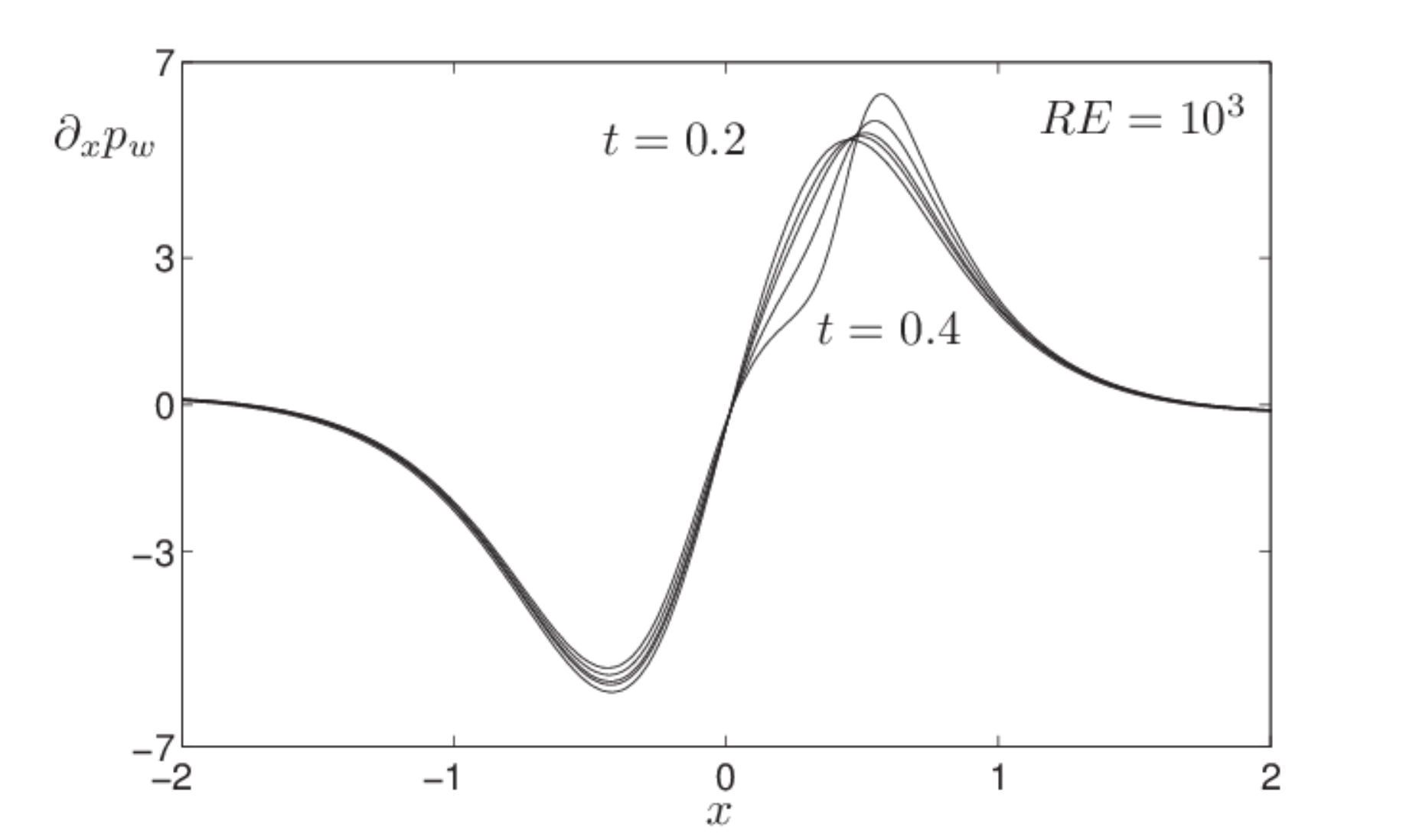}
\end{center}
\caption{Evolution in time, from $t=0.2$ up to $t=0.4$ (with increments of 0.05),
 of the streamwise pressure gradient on the wall. Reynolds number is $Re=10^3$.
 The local change close to the maximum is the result of the large-scale interaction.}
\label{dpdx_a}
\end{figure}

\begin{figure}
\begin{center}
\includegraphics[width=11cm,height=6cm]{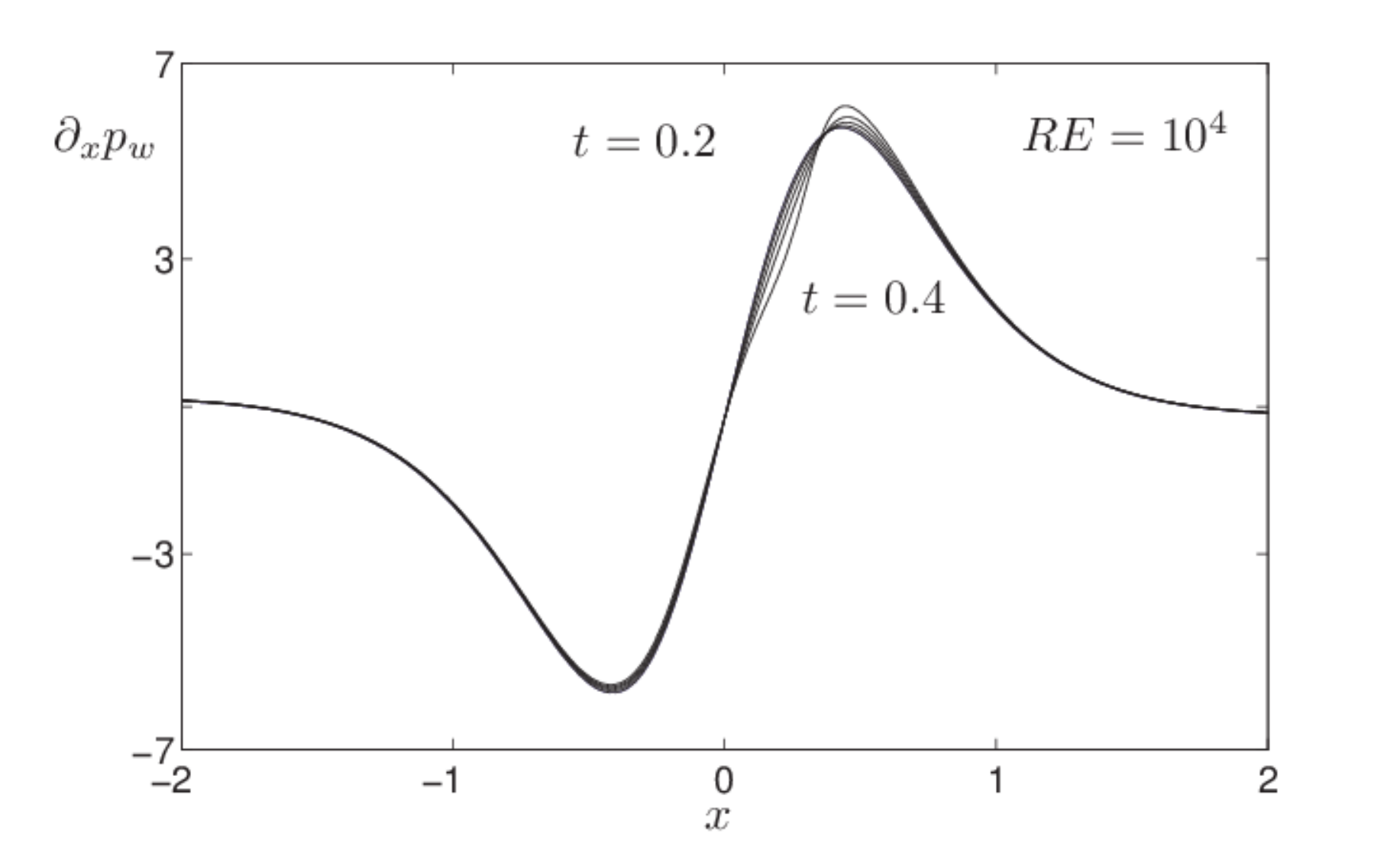}
\end{center}
 \caption{Evolution in time, from $t=0.2$ up to $t=0.4$ (with increments of 0.05),
 of the streamwise pressure gradient on the wall. Reynolds number is $Re=10^4$.
 The local change close to the maximum is the result of the large-scale interaction.}
\label{dpdx_b}
\end{figure}

\begin{figure}
\begin{center}
\includegraphics[width=11cm,height=6cm]{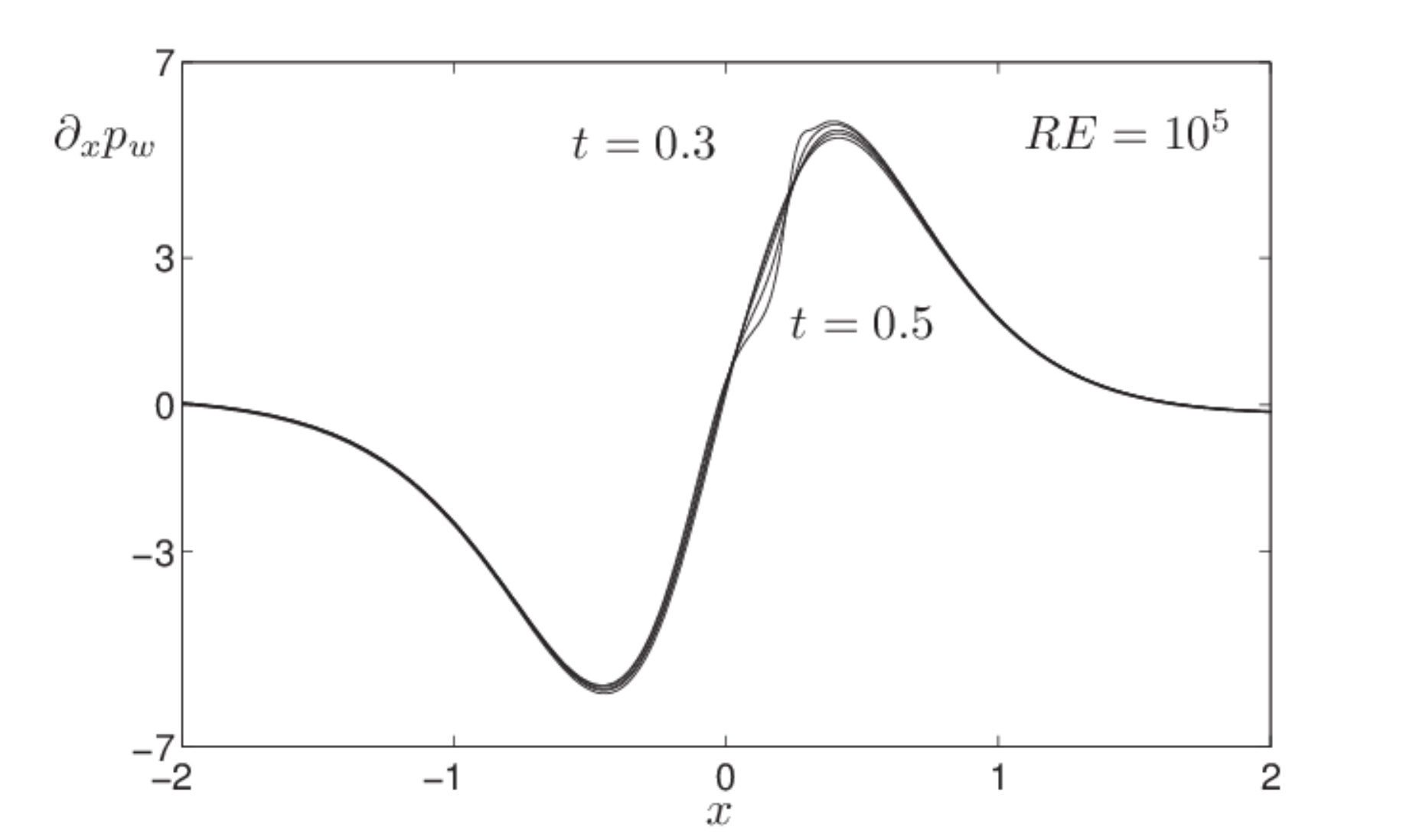}
\end{center}
\caption{Evolution in time, from $t=0.3$ up to $t=0.5$ (with increments of 0.05),
 of the streamwise pressure gradient on the wall. Reynolds number is $Re=10^5$.
 The local change close to the maximum is the result of the large-scale interaction.}
  \label{dpdx_c}
\end{figure}

In particular, given that we are mostly interested in the phenomena occurring
close to the boundary, we shall focus on the pressure gradient at the wall
defined as $\partial_x p_w\equiv \frac{\partial p}{\partial x}|_{y=0}=
-\frac{1}{Re}\frac{\partial{\bar{y}}}{\partial y}\frac{\partial\omega}
{\partial \bar{y}}|_{y=0}$.
The time evolution of $\partial_x p_w$  is shown,
for different $Re$, in Figs. \ref{dpdx_a}, \ref{dpdx_b}, and \ref{dpdx_c}.
As expected, during the early stages of the evolution  the pressure gradient
experiences only small  changes (which are due to the effect of  the viscosity),
that do not have any remarkable effect on the flow dynamics.
However, at a later time (different for different $Re$ numbers) one can observe,
close to the maxima, strong variations in the pressure gradients as
the result of the first viscous-inviscid interaction.
The zone where this interaction acts has streamwise spatial dimension comparable with
the characteristic length (the distance of the vortex from the wall) and with the size
of the recirculation region, and therefore we call this interaction (as in
\cite{Cas00,OC02}) the large-scale interaction.

\begin{figure}
\begin{center}
\includegraphics[width=12cm,height=6.5cm]{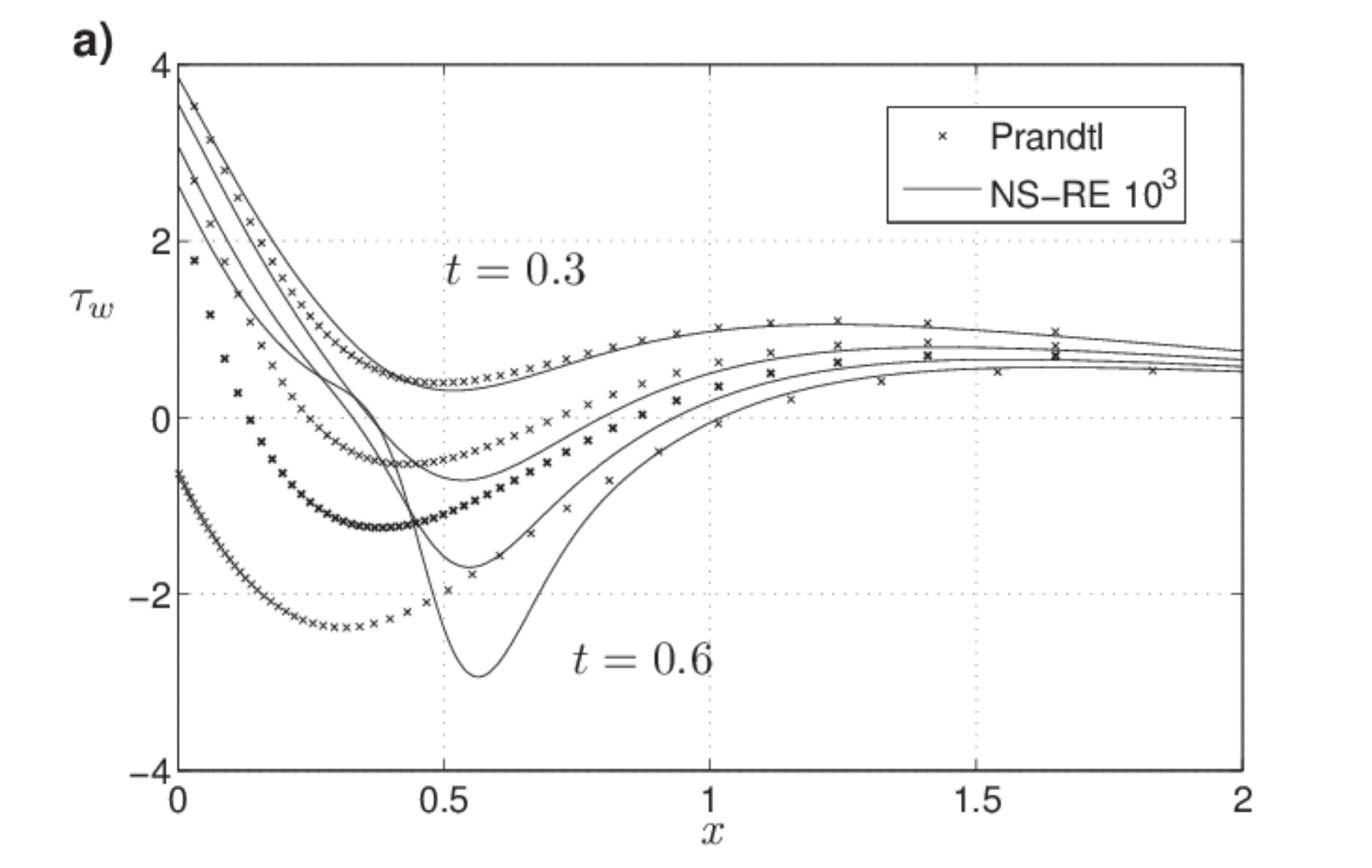}

\includegraphics[width=12cm,height=6.5cm]{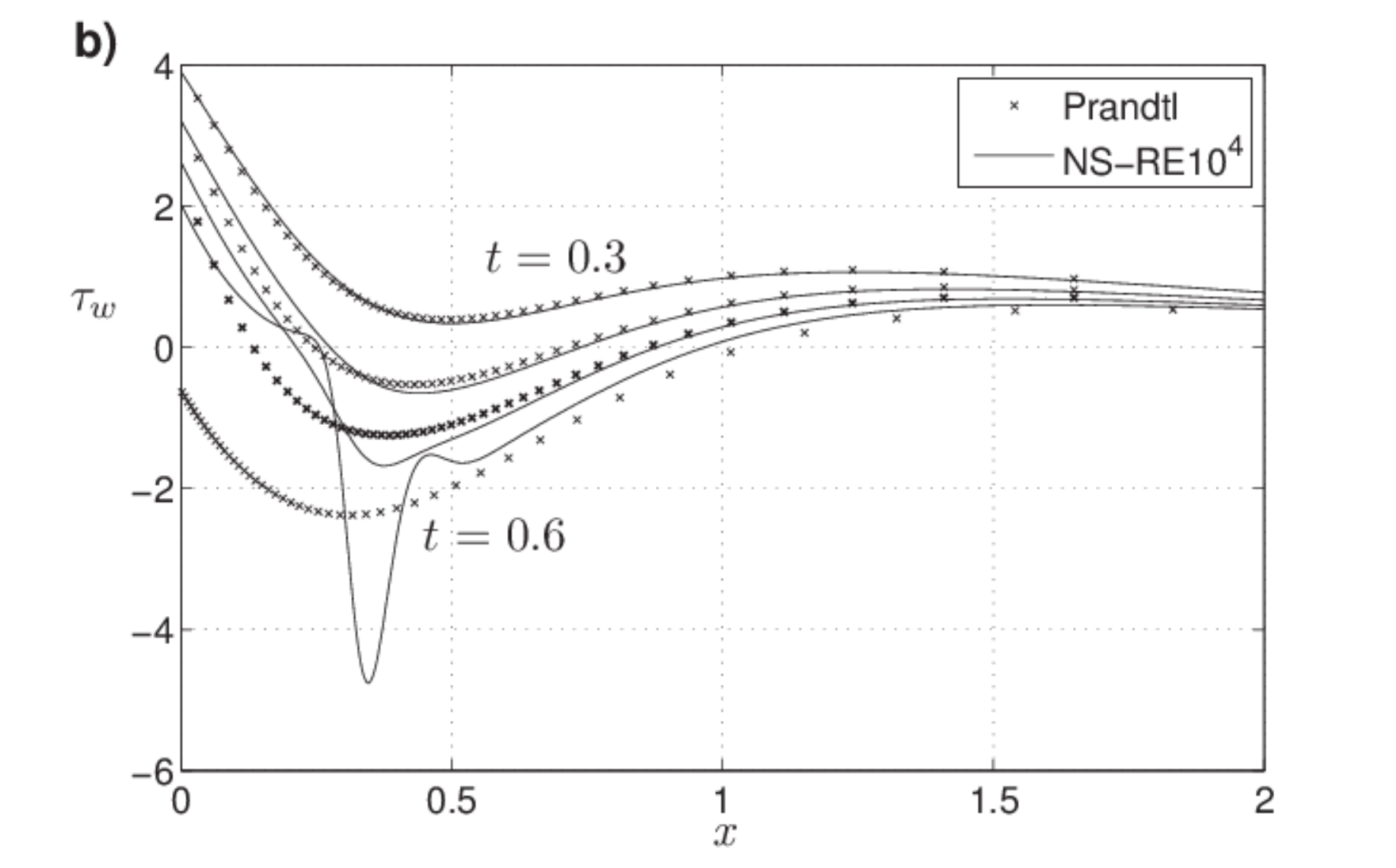}

\includegraphics[width=12cm,height=6.5cm]{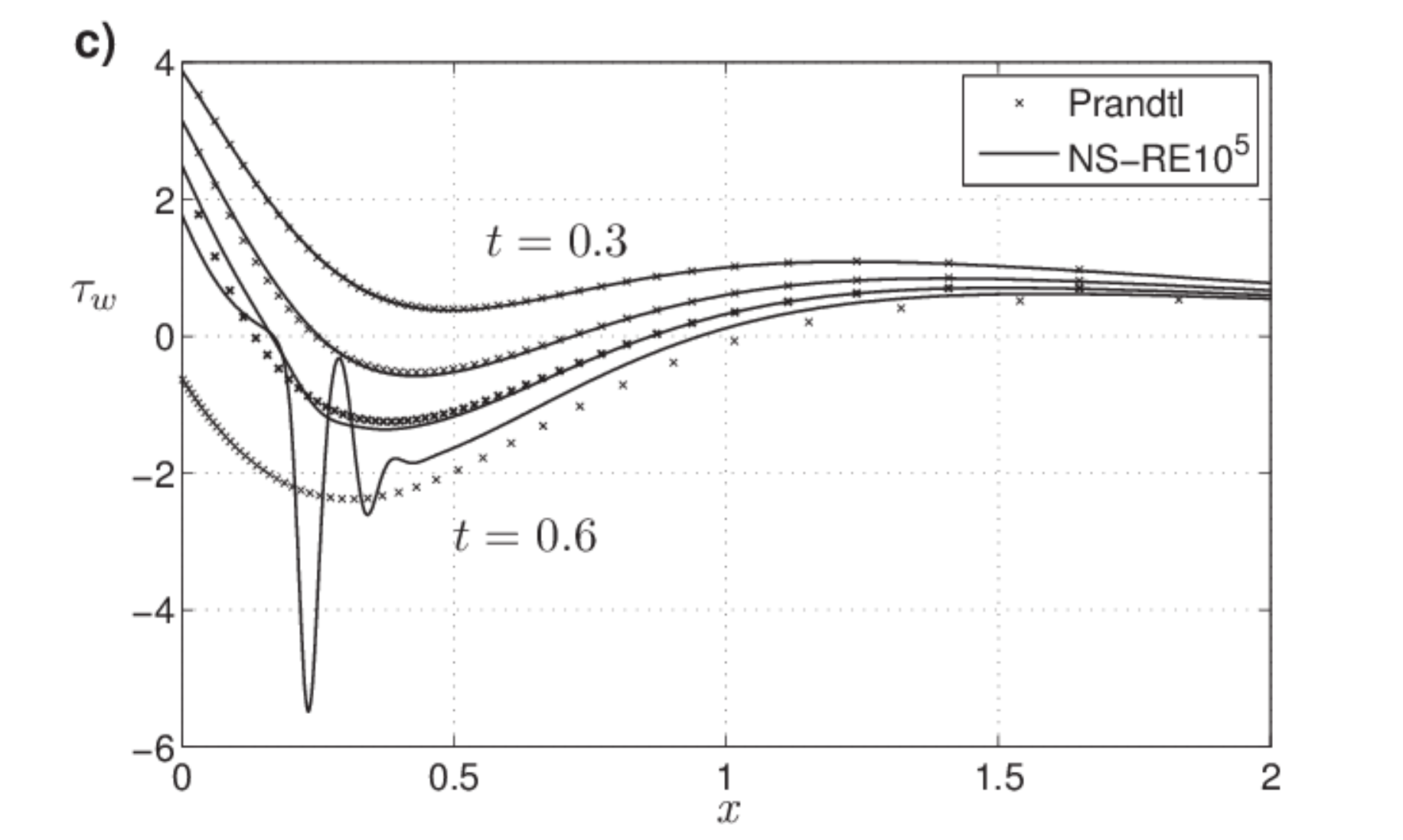}
\end{center}
\caption{A comparison in time between Prandtl and Navier-Stokes  wall shear for
various $Re$ numbers .
The comparison starts at time $t=0.3$ up to time $t=0.6$ (increments of 0.1).
They compare well up to when the large-scale interaction begins, which happens at
$t=0.34,0.38,0.47$ for $Re=10^3,10^4,10^5$ respectively.}
\label{wallshear_comp}
\end{figure}

It is not easy to define the precise time when this large-scale interaction begins;
in fact, also because the interaction occurs over a wide streamwise scale, it is
hard to understand when the variation in $\partial_x p_w$ has remarkable effects
on the flow motion.
However one can observe that, in the temporal range during which $\partial_x p_w$
begins to be visibly different from Prandtl's streamwise pressure gradient,
there is the formation of a pair of inflection points between the maximum and the
inflection point located in $(0,0)$.
These two inflection points are the consequence of the vanishing of the second
derivative of $\partial_x p_w$ and forms at time $t\approx0.34$ for $Re=10^3$, $t\approx0.38$ for $Re=10^4$ and $t\approx0.47$ for $Re=10^5$,
at the spatial locations  $x\approx0.29$, $x\approx0.2$, $x\approx0.17$ respectively.
These inflection-points carry physical meaning, as they are the precursor of
the formation of local minima in the pressure gradient which will eventually become
negative (see e.g. Fig.\ref{stream1000_1}b of the next Section for the case $Re=10^3$).
These local negative minima in $\partial_x p_w$ reflects a  pressure gradient adverse
to the flow motion in the primary recirculation region near the wall,
which forces the formation of a secondary recirculation region,
and introduces a change in the qualitative behavior of the flow.
For this reason we define the time when large-scale interaction begins as the
time when these inflection point forms in  $\partial_{x}p_{w}$.
At this time one can observe significant quantitative differences between Prandtl's
and Navier-Stokes solutions.
In Fig.\ref{wallshear_comp} we show the behavior of the wall shear stress
(which is basically the vorticity at the wall) for Prandtl's solution
and Navier-Stokes solutions for different $Re$.
The most visible differences are evident during the large-scale interaction
stage in all cases.
For $Re=10^3$ (Fig.\ref{wallshear_comp}a) Navier-Stokes and Prandtl's wall shear
are very close up to time $t=0.3$,  reflecting the good agreement
between the two solutions in the whole boundary layer.
On the other hand at time $t=0.4$ one can already see significant discrepancies,
and these are the consequence of the large-scale interaction influencing
Navier-Stokes  solution.
This influence is also shown for $Re=10^4, 10^5$ in Figs.\ref{wallshear_comp}b,c
at $t=0.4$ and $t=0.5$ respectively.

\begin{figure}
\begin{center}
\includegraphics[width=10cm,height=6cm]{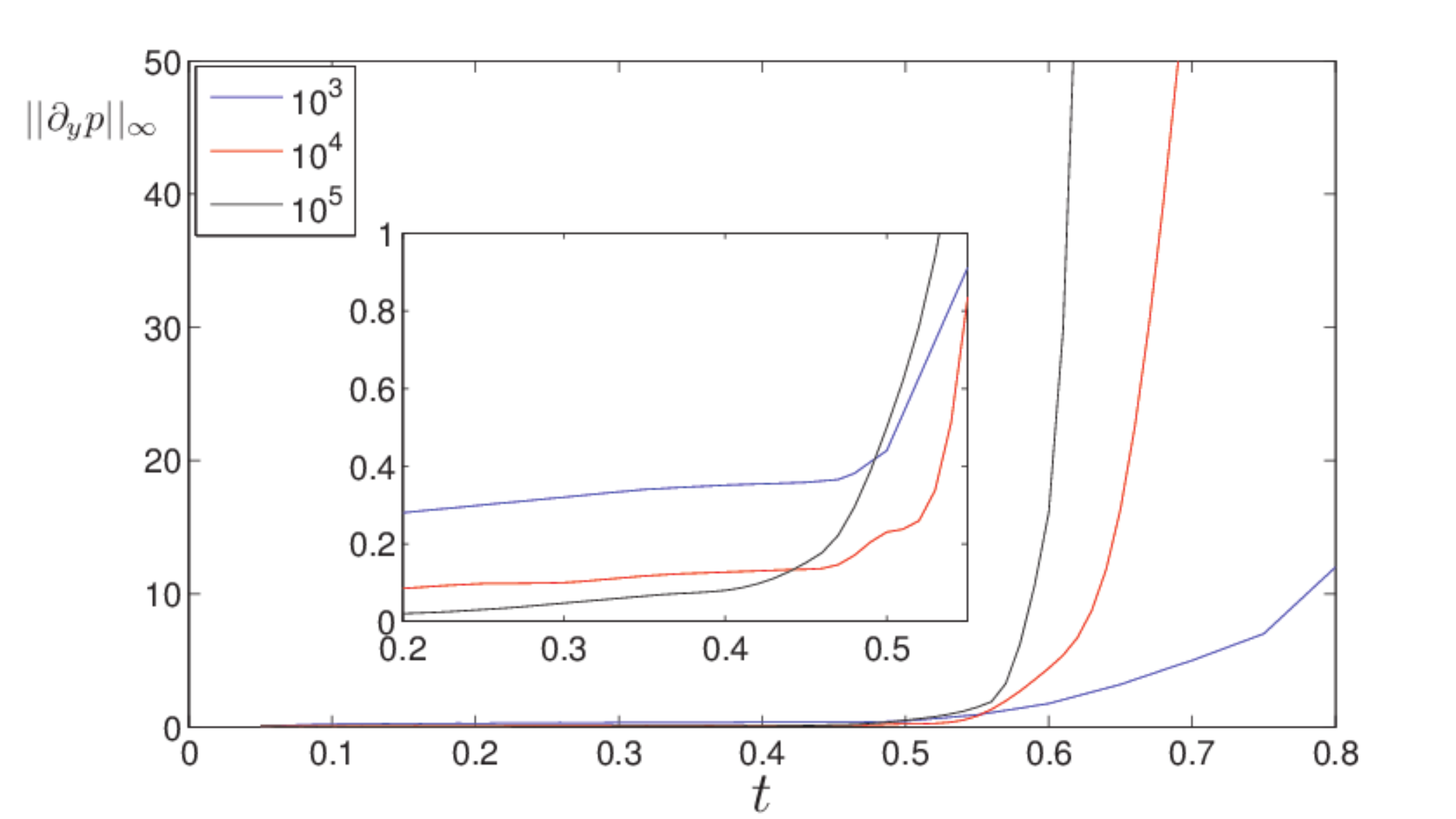}
\end{center}
\caption{The evolution in time of the sup norm of $\partial_y p$ in the boundary
layer from time 0.2 to 0.8.
In the inset the evolution from 0.2 to 0.6.
Note how $Re^{-1/2}\partial_y p$ remains of order $O(Re^{-1/2})$ during the whole
large-scale interaction stage.}
\label{dpdy}
\end{figure}

We conclude this Section stressing how the discrepancies between
the Navier-Stokes and Prandtl's solutions
during  this stage are merely quantitative, while the overall qualitative
properties of the two flows are quite similar.
In fact during the large-scale interaction only one big  recirculation region
is present in the NS flow (likewise in classical BLT).
Moreover the magnitude of the normal pressure gradient re-scaled according to the
boundary layer variable ($\partial_Y p=Re^{-1/2}\partial_y p$)
is of order $O(Re^{-1/2})$.
This is compatible with the Boundary Layer assumption predicting that the normal
pressure gradient is always zero.
This is shown in Fig.\ref{dpdy} where one can see the evolution in time of
$||\partial_y p||_{\infty}$ evaluated inside the boundary layer.

It is therefore clear that the large scale interaction has a different character
compared with the viscous-inviscid interaction visible in Prandtl's flow.
The latter, in fact, which  is signaled by the rapidly growth of the spike in the streamlines, begins at the time $t_k=0.85$ while, at least for the $Re$ considered
here, the large-scale interaction begins at times between  $0.3$ and $0.5$.
More importantly, during the large-scale interaction  no large gradient is present
and the flow remains confined in the boundary layer without eruption toward
the outer flow.
These phenomena characterize the following stage of the evolution and will be
analyzed in the next Section.

\section{Navier-Stokes results: Small-Scale Interaction}
\setcounter{equation}{0}

The characteristics of the large-scale interaction bear no resemblance with the
viscous-inviscid interaction developed by Prandtl's solution which is characterized
by the formation of a spike in the streamlines and vorticity contours.
However the large-scale interaction in Navier-Stokes solutions is the precursor
of another interaction, acting on a smaller scale.
We shall see that this phenomenon occurs only for moderate-high $Re$
(i.e. $10^4\leq Re\leq 10^5$) numbers.

\subsection{Moderate-high Reynolds numbers: $10^4\leq Re\leq 10^5$}

\begin{figure}[h]
\begin{center}
\hspace*{-3cm}
\includegraphics[width=20cm,height=15.5cm]{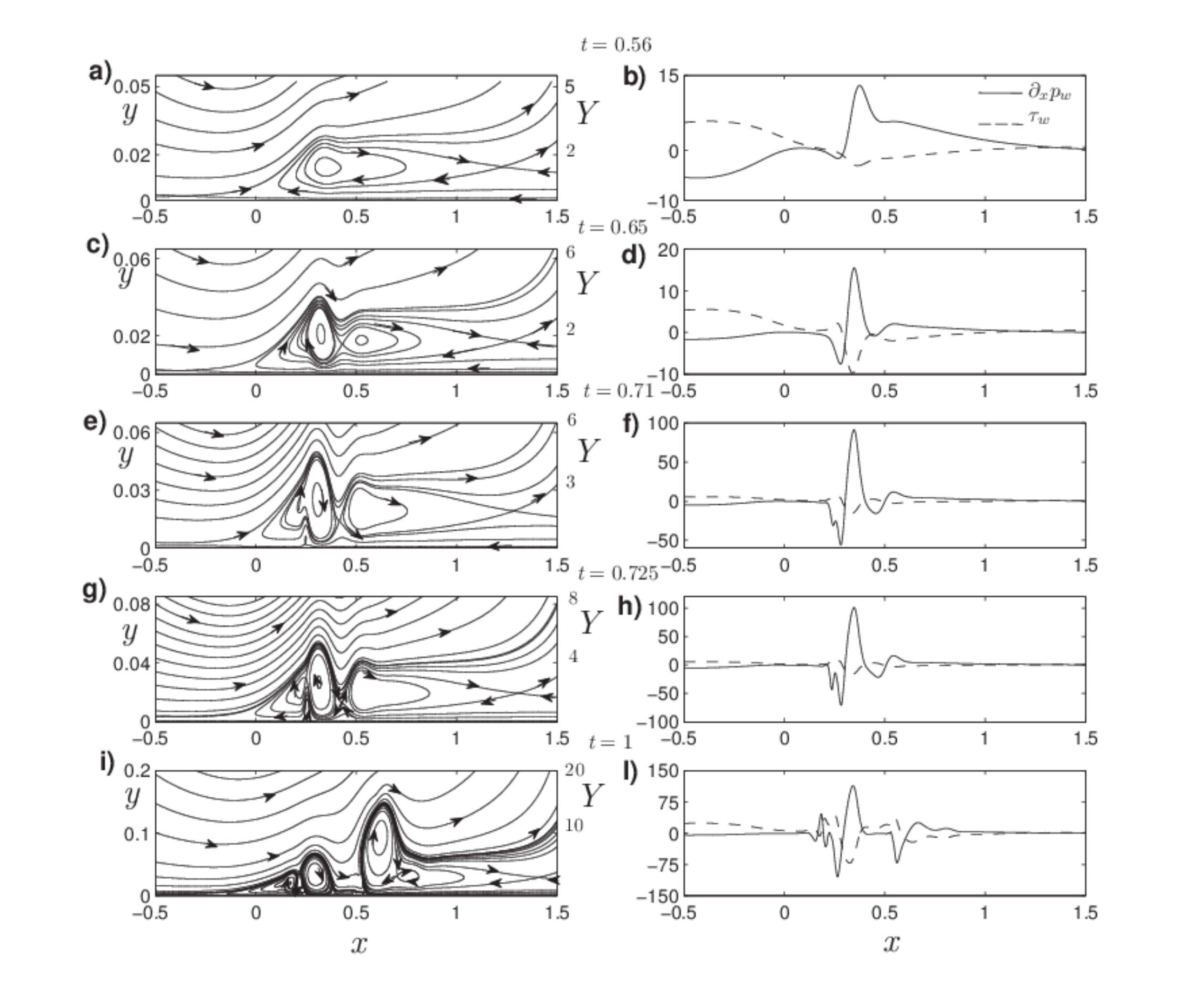}
\end{center}
\caption{ $Re=10^4$. Streamlines on the left, wall shear stress (dotted) and streamwise pressure
gradient on the wall (dashed) on the right.
At $t=0.56$ a kink is visible above and to the left of the recirculation region.
This kink corresponds to the formation of a negative minimum (upstream respect to the near wall flow
motion) and a positive maximum (downstream) of the streamwise pressure gradient.
The splitting of the recirculation region, caused by the presence of alternating
adverse-favorable-adverse pressure gradients, is visible at $t=0.65$.
A new recirculation region is visible at $t=0.71$ due to the adverse pressure gradient created
by the minimum to the left of the main maximum.
At $t=0.725$ this new recirculation region causes a further splitting of the primary recirculation region.
At $t=1$ several eddies are visible and the splitting of the recirculation regions continues. }
\label{stream10000}
\end{figure}

To describe this new interaction in Fig.\ref{stream10000}a we show, for $Re=10^4$,
the streamlines  at time $t=0.56$, while in Fig.\ref{stream10000}b
we show the wall shear $\tau_w$ (dotted) and the streamwise pressure gradients on the
wall $\partial_{x}p_{w}$ (dashed).
In the streamlines it is clearly visible the formation of a kink located
above and to the left of the recirculation region; in correspondence one can observe
a strong streamwise variation in $\tau_w$ and $\partial_{x}p_{w}$ and the rapid alternation of critical points.
The first minimum of $\partial_{x}p_{w}$ forms on the left of the main maximum
(the time at which the minimum appears is roughly $t\approx0.52$) at
$x\approx0.21$, close to the streamwise location
where we observed (in the previous Section) the change of concavity  that
characterizes the large-scale interaction.
At time $t\approx0.56$ this minimum becomes negative; one therefore has a
pressure gradient that is adverse with respect to the flow motion of the
recirculation region close the boundary.
In Fig.\ref{stream10000}b one observes that the minimum in the pressure
gradient is followed, through a sharp increase, by a positive maximum (where
the pressure gradient is therefore favorable).
The fluid portion between the minimum and the maximum is therefore
strongly compressed in the streamwise direction.
This compression accelerates the evolution of the kink (visible
in Fig.\ref{stream10000}a)  in a spike (visible in Fig.\ref{stream10000}c),
and it is the responsible for the growth in the normal direction of the boundary
layer, and of the subsequent vorticity eruption.
The presence of the kink and of the subsequent spike signals a new kind of interaction
between the BL and the outer flow, which is called small scale interaction, \cite{OC02}.

Notice also, in Fig.\ref{stream10000}b, the presence of a second local minimum
which forms at $x\approx 0.49$.
This secondary minimum will soon become negative (see Fig.\ref{stream10000}d),
with the consequent presence of a further adverse pressure gradient; this
leads to the splitting of the recirculation region in two recirculation
regions which is clearly visible in  Fig.\ref{stream10000}c.
On the other hand the primary minimum in the pressure gradient, which in
Fig.\ref{stream10000}d is located at $x\approx 0.28$, creates an adverse
pressure gradient: this creates the conditions for the birth of a further recirculation
region  attached to the boundary, which is visible in Fig.\ref{stream10000}e.

The newly created recirculation region, pushing the fluid from below, causes
a further splitting of the primary recirculation region, visible in
Fig.\ref{stream10000}g.
The process of creation of vortical structures continues over and over, while the structure of the pressure gradient becomes more and more complicated.
This can be seen in Figs.\ref{stream10000}i-l.

\begin{figure}[h]
\begin{center}
\hspace*{-3cm}
\includegraphics[width=20cm,height=12.5cm]{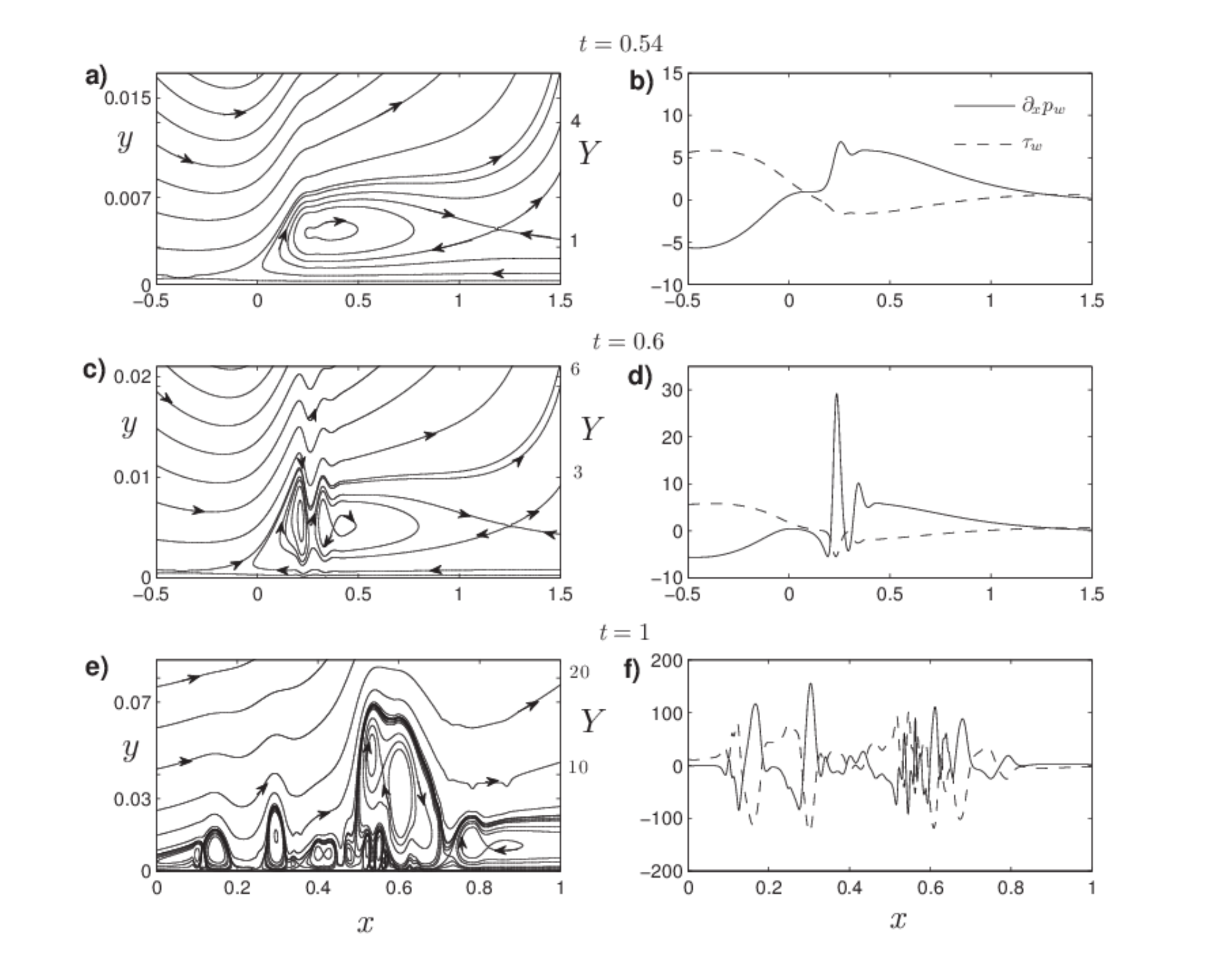}
\end{center}
\caption{ $Re=10^5$. Streamlines on the left, wall shear stress (dotted) and streamwise pressure gradient on the wall (dashed) on the right.
The kink on the streamlines visible at $t=0.54$ is the result of the formation of an upstream and downstream minimum in $\partial_x p_w$. At $t=0.6$ the recirculation region is split caused by the presence of alternating adverse-favorable-adverse pressure gradients. At $t=1$  complicated structures are present in the flow. }
\label{stream100000}
\end{figure}

The behavior for $Re=10^5$ is qualitatively similar the main difference being that
the whole sequence of formation of alternating critical points in the pressure gradient
followed by the splitting of the recirculation region is considerably faster.
In fact the first downstream minimum in the pressure gradient  forms
at $t\approx0.52$ in $x\approx0.3$, while the second upstream minimum forms
at time $t\approx0.53$ in $x\approx0.13$.
All this is clearly visible in Fig.\ref{stream100000}.

\subsection{Low Reynolds numbers: $Re\simeq 10^3$}
\begin{figure}
\begin{center}
\includegraphics[width=13cm,height=6.5cm]{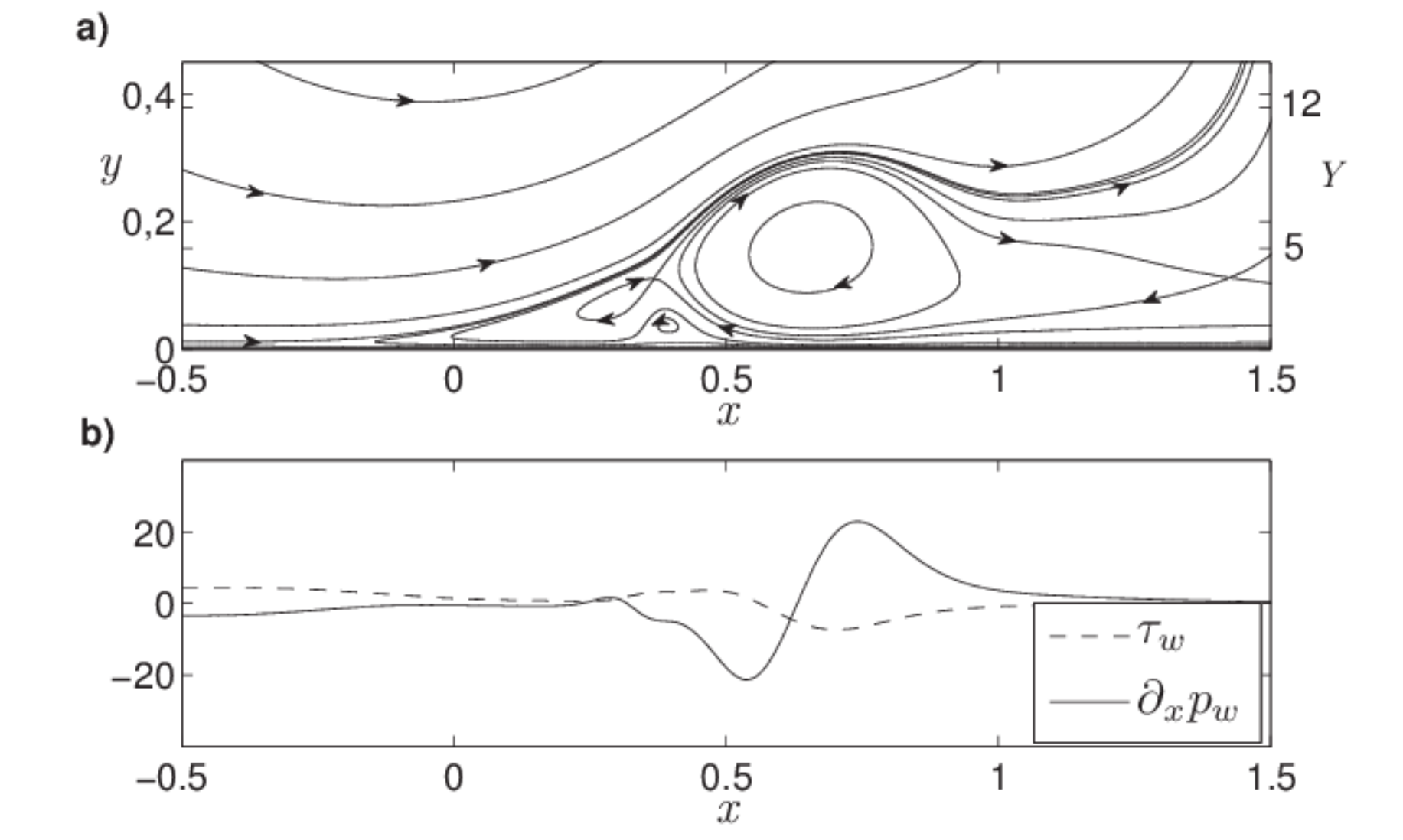}
\end{center}
\caption{$Re=10^3$, t=1.02. a) streamlines. b) wall shear (dotted), streamwise pressure gradient at the wall (dashed). The primary recirculation region splits in two co-rotating eddies, and a new recirculation region forms underneath soon after $t=1$. }
\label{stream1000_1}
\end{figure}

For $Re=10^3$ the flow evolution is quite different from what we have seen
for $Re=10^4-10^5$.
In fact after the large-scale interaction stage only the upstream minimum
forms at $t\approx0.49$ in $x\approx0.28$; this minimum becomes negative at time
$t\approx0.54$ in $x\approx0.35$.
The splitting of the recirculation region happens at $t=0.95$ on the left side of
the primary recirculation region, and a secondary recirculation forms underneath
the primary soon after $t=1$ as consequence of the adverse pressure
gradient to the recirculation flow motion.
In Fig.\ref{stream1000_1}  we show the streamlines, the wall shear
and the streamwise pressure gradient at the wall  at time $t=1.02$:
no formation of any local minimum downstream to the right of the local maximum
is present, no kink or spike  in the streamlines, and no large gradients in
$\partial_{x}p_{w}$ or $\tau_w$.
Therefore the typical characterizations of the small-scale interaction are not
present for $Re=10^3$.
We have checked that, up to time $t=6$, no small-scale interaction can be detected.

The different flow evolution observed for $Re=10^3$ can be explained with the
larger diffusive effects acting for lower $Re$ number, which prevents
the streamwise compression that, for $Re=10^4-10^5$ led to the formation of the spike.
This difference can be seen also in terms of pressure gradient; in fact no minimum in
$\partial_x p_w$ forms downstream and the pressure gradient keeps the rather simple
structure visible in Fig.\ref{stream1000_1}b.
Clearly the formation of the local downstream minimum is a crucial event in the
flow evolution, as it causes an early splitting of the recirculation region and
initiates the cascade of eddies that characterizes the moderate-high $Re$ regime.

The boundary between the low Reynolds number regime and the moderate-high
regime is in between $Re=2\cdot 10^3$ (for this $Re$ we have not detected the
small scale interaction) and  $Re=3\cdot 10^3$ (for which, on the other hand,
the small-scale regime is visible).

\section{Separation, dipolar structures and vorticity production}
The description of the unsteady separation of  the previous sections was based
on the analysis of the evolution of the streamwise pressure gradient and of the vorticity at the wall.
In this Section we shall look at  the boundary layer dynamics from a different
perspective.
Namely we shall see how, for moderate-high $Re$, an important event occurring
during the separation process, is the creation of several vortex-dipoles, and that
the reciprocal interaction between these structures leads to a sharp increase
in the enstrophy of the flow.

The equations for the evolution of the energy and of the
enstrophy of the flow within the boundary layer $D$ (see Section 3.2 for the definition of $D$)
write as:
\begin{eqnarray}
\frac{dE(t)}{dt}&=&
-\frac{1}{Re}\Omega(t)+I^\omega(t) + NT_1\label{energy}\\
\frac{d\Omega(t)}{dt}&=&-\frac{2}{Re}P(t)+
2I^p(t)+NT_2 \label{enstrophy}
\end{eqnarray}
where
\begin{eqnarray}
I^{\omega}(t)&=&-\frac{1}{Re} \int_{-\infty}^{+\infty}u_{y=0}\cdot\omega_{|y=0}dx \, ,  \nonumber \\
I^p(t)&=& \int_{-\infty}^{+\infty}\omega_{|y=0}\cdot\partial_x p_w
dx \, ,\nonumber
\end{eqnarray}
and
\begin{eqnarray}
NT_1&=&{-\frac{1}{2}\int_{\partial D}\mathbf{u}^2(\mathbf{u}\cdot\mathbf{n}) dl} -\int_{\partial D}p(\mathbf{u}\cdot\mathbf{n}) dx
-\frac{1}{Re}\int_{-\infty}^\infty (\omega u)|_{y=Y_{BL}} dx \; , \nonumber \\
NT_2&=&\frac{2}{Re} \int_{-\infty}^{+\infty}\left(\omega\cdot \partial_y\omega\right)_{|y=Y_{BL}}
dx \; , \nonumber
\end{eqnarray}

being $\mathbf{n}$ the exterior normal to $\partial D$.
The $NT_i$ terms are negligible because at $y=Y_{BL}$ the vorticity $\omega$ is very small
and the normal component of the velocity is very close to be an odd function while $\mathbf{u}^2$
and $p$ are even.
We shall not consider these terms in the rest of our analysis.

The energy decreases, as the negative term due to the enstrophy is larger than
$I^{\omega}(t)$ (see Figs.\ref{enstrophyfig}b-d).
From \eqref{enstrophy} one can see that the only way for the enstrophy
to increase is via the integral term $I^p$
which is related to the vorticity and to the vorticity flux at the boundary.
We shall see that the enstrophy within the boundary layer can in fact increase,
and  how the evolution in time of $\Omega(t)$ is related to important events
characterizing the separation process.

\subsection{Large-scale interaction: the detachment of boundary layer}
\begin{figure}
\centering
\subfigure[$Re=10^3,t=.65$]{\includegraphics[width=6.0cm,height=3.5cm]{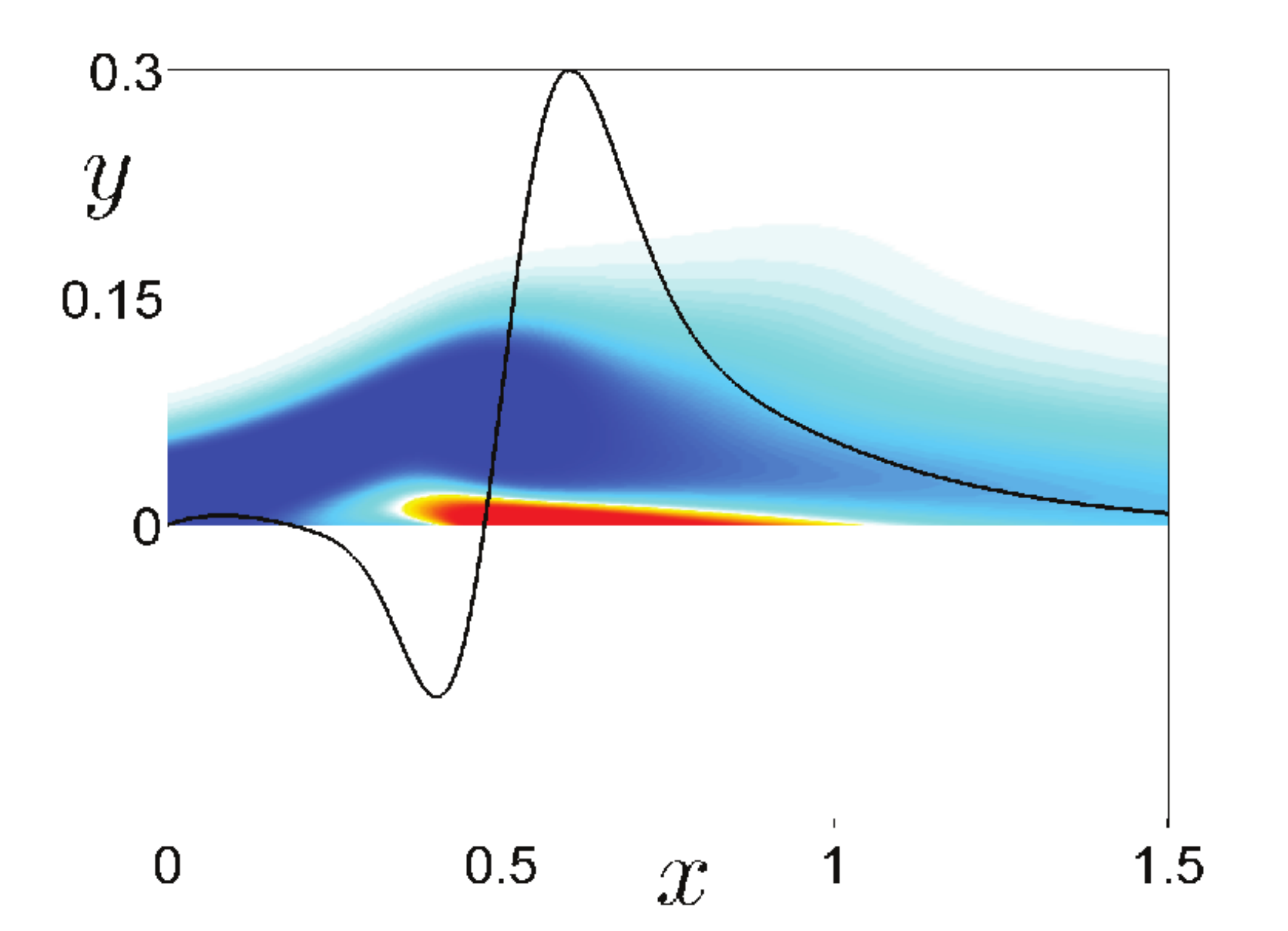}}
\subfigure[$Re=10^3,t=.80$]{\includegraphics[width=6.0cm,height=3.5cm]{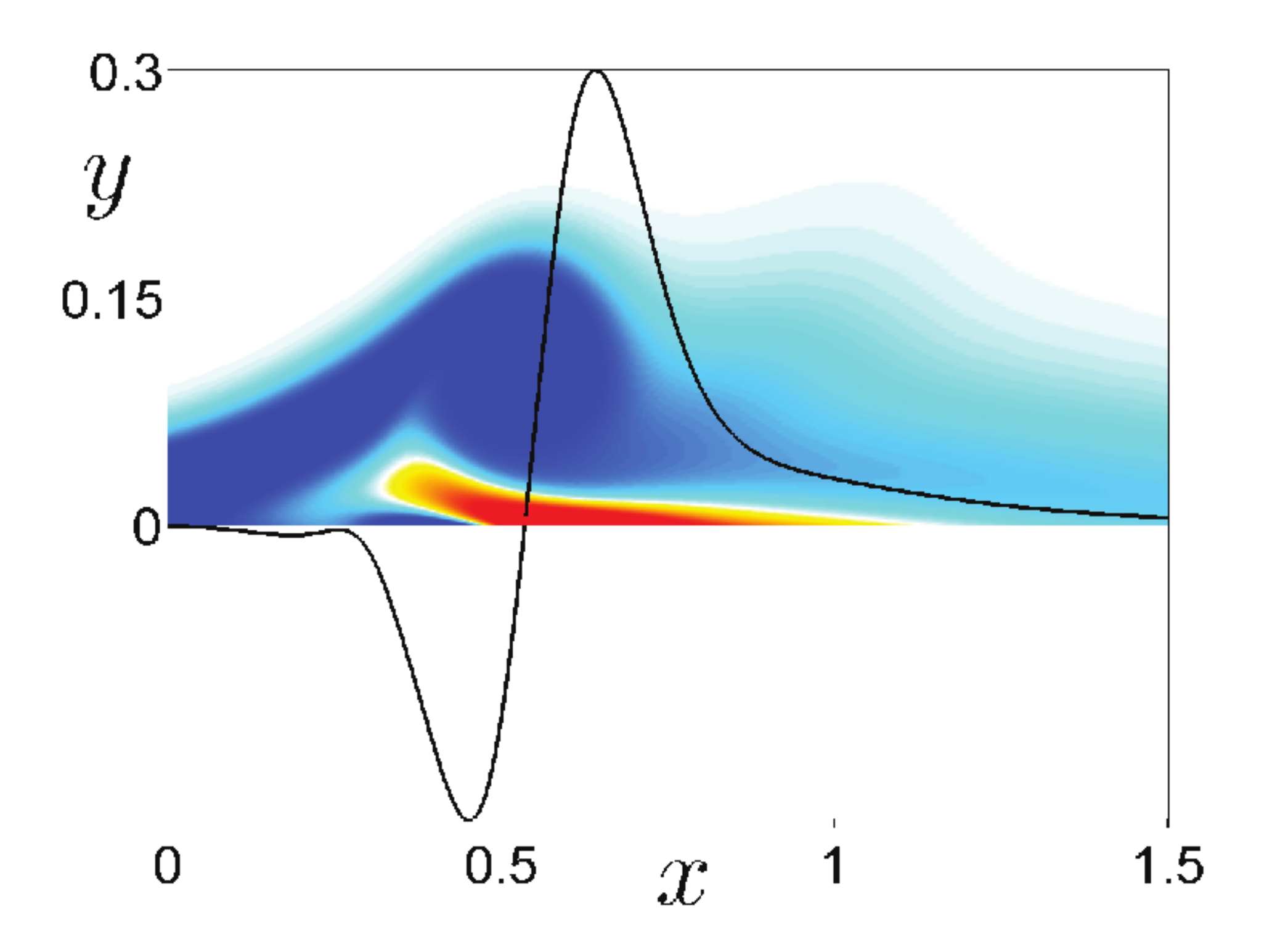}}
\subfigure[$Re=10^3,t=1.0$]{\includegraphics[width=6.0cm,height=3.5cm]{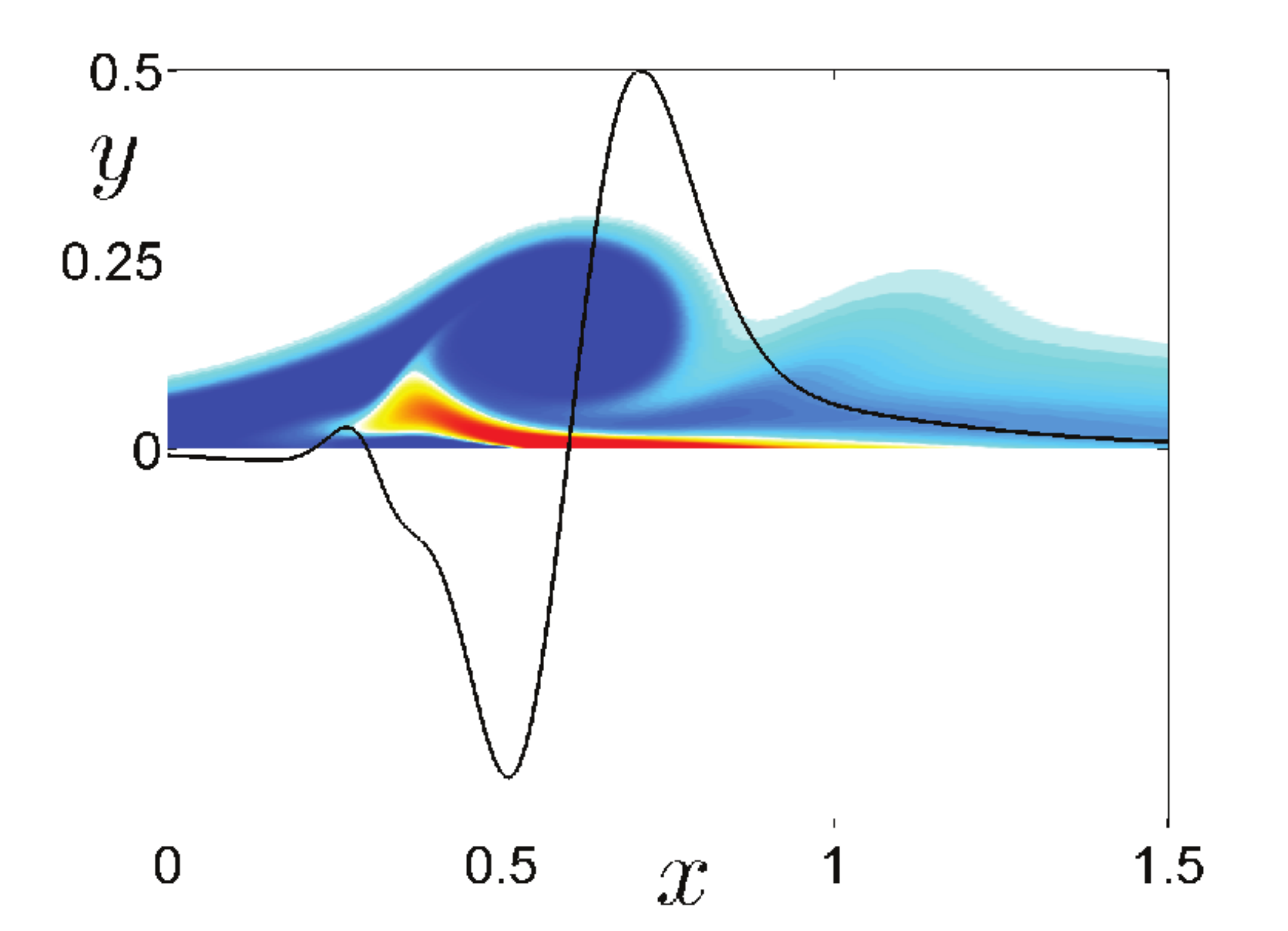}}
\subfigure[$Re=10^3,t=1.5$]{\includegraphics[width=6.0cm,height=3.5cm]{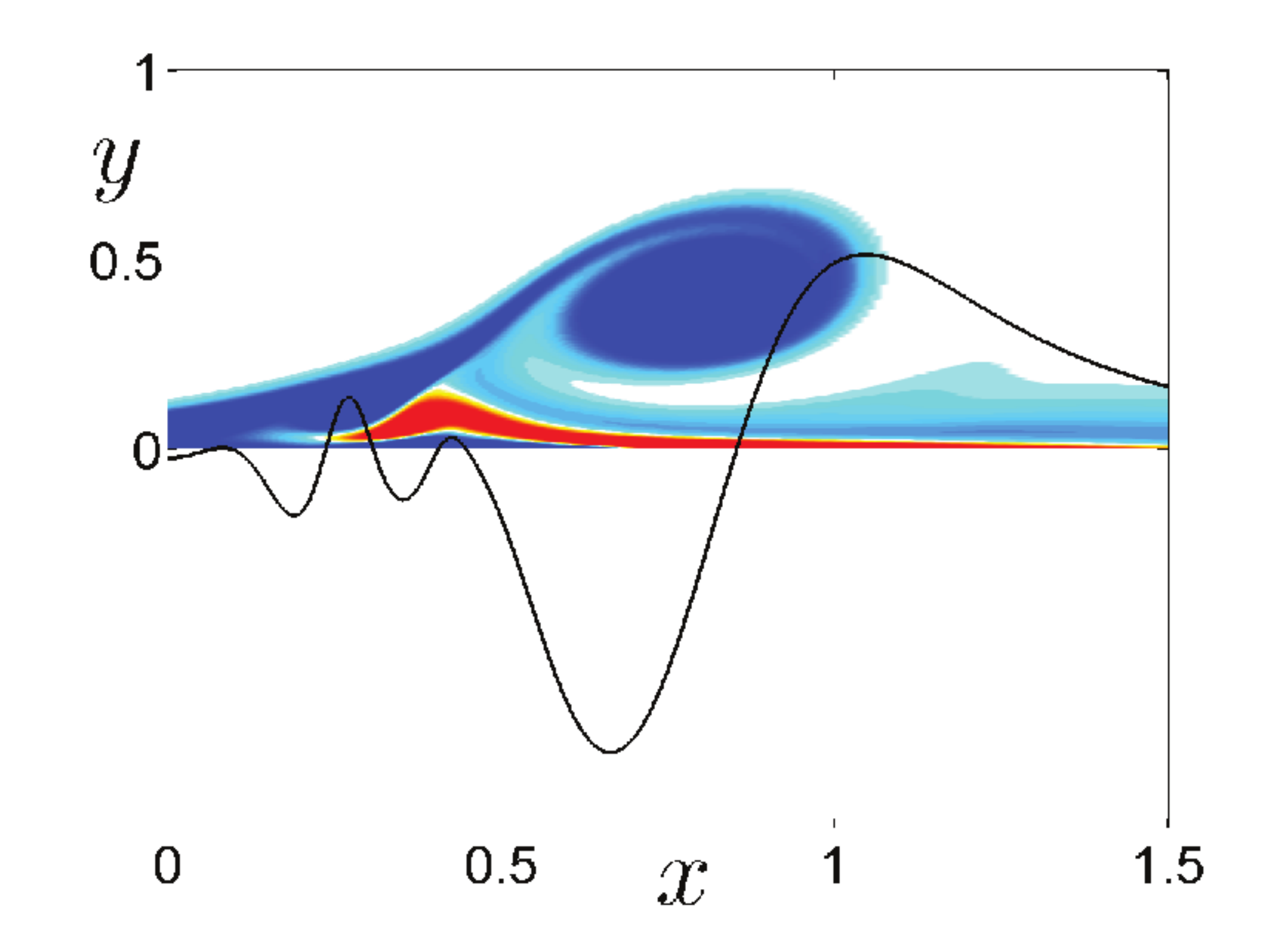}}
\caption{Vorticity contour levels for $Re=10^3$ at different times compared
with $\partial_x p_w$ (rescaled to fit the normal extension of vorticity). The
blue colors represents negative vorticity, red/yellow positive
vorticity. }
\label{vorticity1000}
\end{figure}

\indent
Prior to the large-scale interaction the flow evolution is almost
the same for all the $Re$ numbers.
The {\it no-slip} condition at the wall stops the flow motion and creates a boundary
layer of negative vorticity  $BL_{-}$.
At $t\approx 0.33$ the adverse pressure gradient imposed by the primary
vortex leads  to the formation of a positive vorticity zone
$b_{1+}$ under $BL_-$.
This situation is visible in Figs.\ref{vorticity1000} for $Re=10^3$, where the
positive vorticity $b_{1+}$ still remains beneath $BL_-$ at $t=1$ and $t=1.5$.
As time passes, the part of $BL_-$ located to the right of the primary vortex,
rolls-up and moves to the right due to the velocity
imposed by the primary vortex itself, being also pushed up by the
positive vorticity zone $b_{1+}$.
As $BL_{-}$ detaches from the wall, it creates a clockwise rotation
close to the wall, and therefore the flow particles, especially those  in
$b_{1+}$, are accelerated from right to left.
Therefore  the streamwise pressure gradient becomes negative
on the left of the core of $BL_-$, creating an adverse pressure gradient to the
recirculation region, see for example Fig.\ref{stream10000}.
This variation in the streamwise pressure gradient is gradual, and it is a direct
consequence of the detachment process of the boundary layer $BL_-$, which begins to
interact with the inviscid outer flow leading to the large--scale interaction stage
described in Section 5.
\begin{figure}
\centering
\subfigure[$Re=10^4,t=0.7$]{\includegraphics[width=6cm,height=3.5cm]{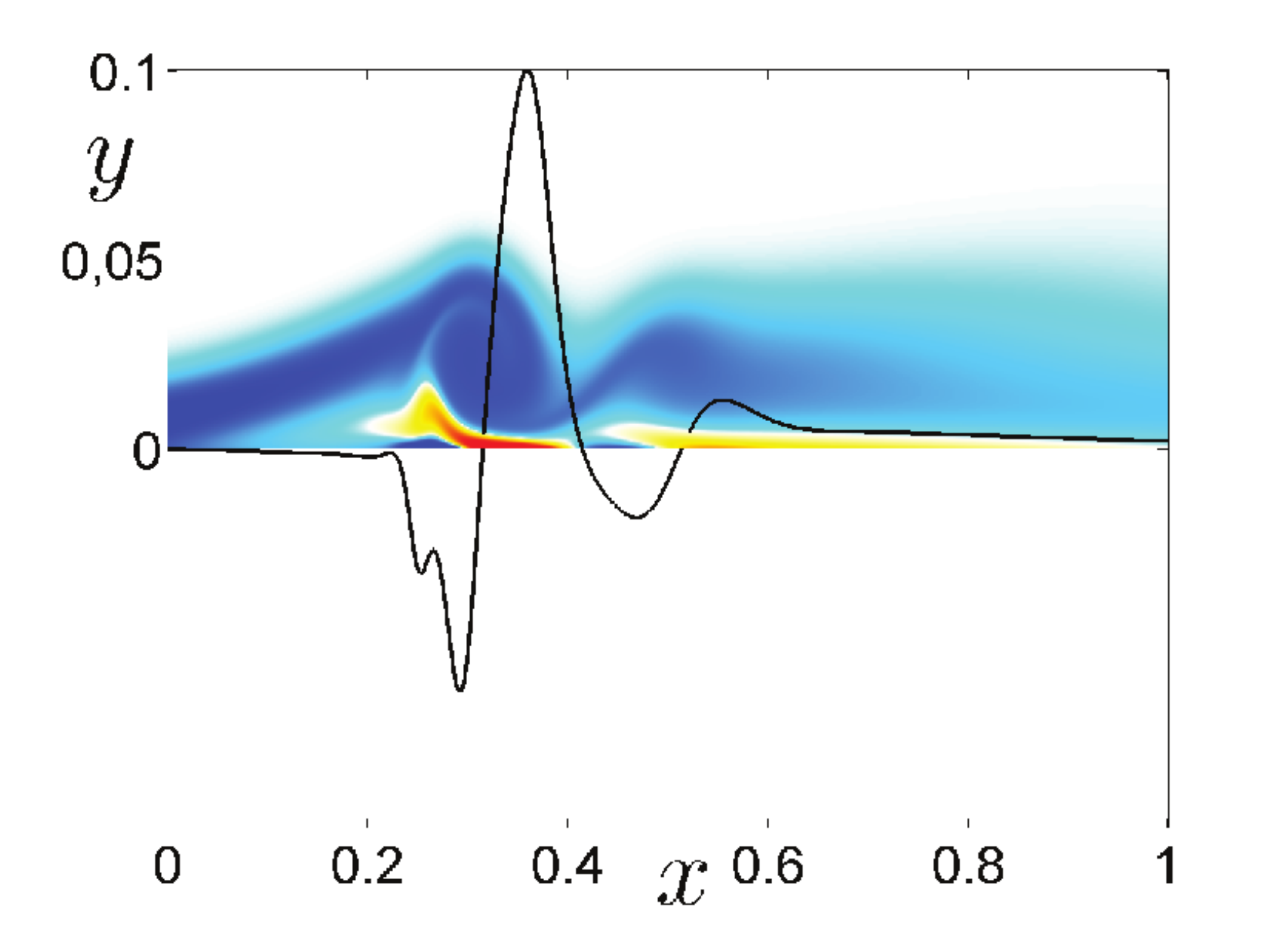}}
\subfigure[$Re=10^4,t=0.78$]{\includegraphics[width=6cm,height=3.5cm]{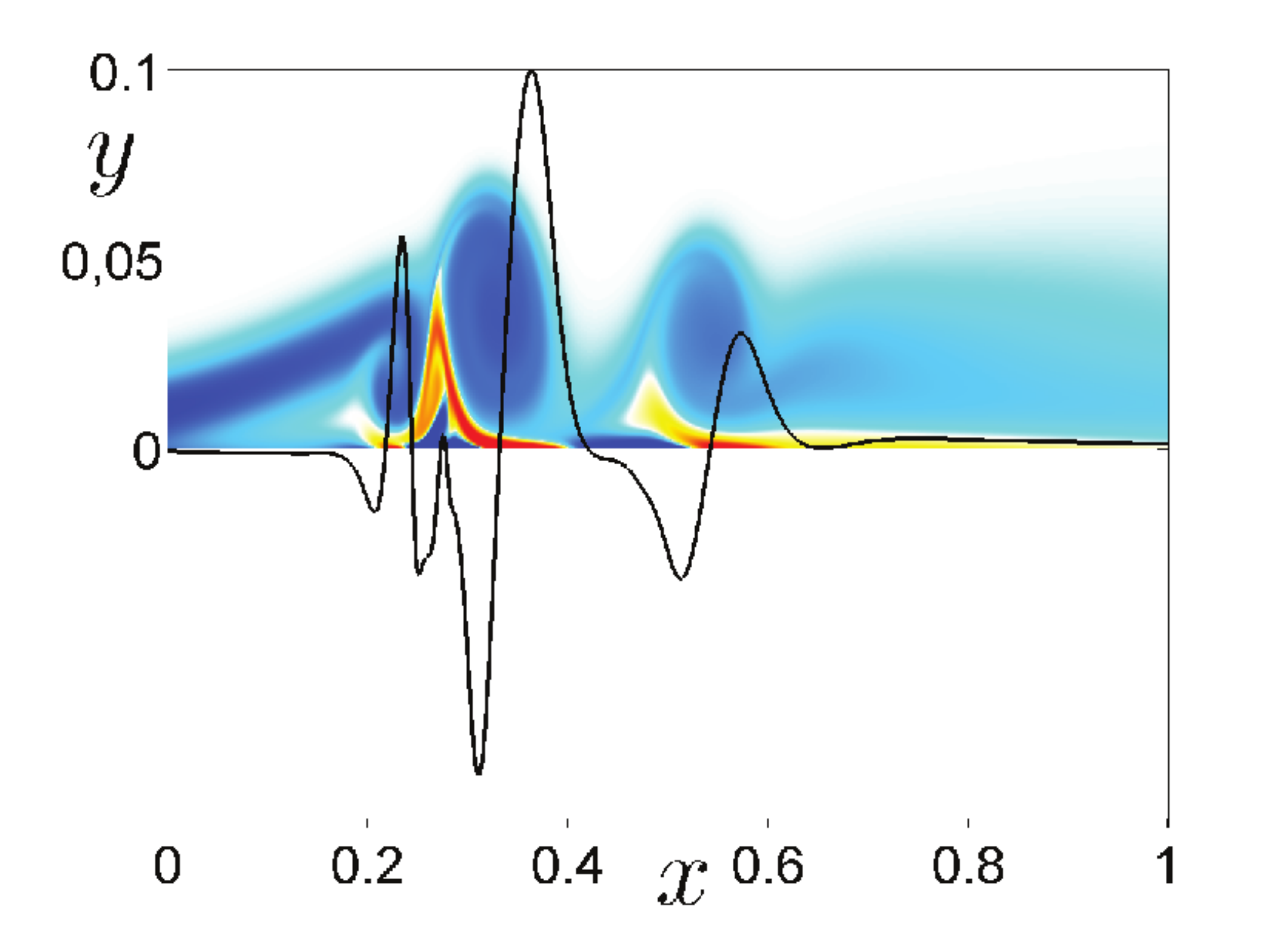}}
\subfigure[$Re=10^4,t=0.9$]{\includegraphics[width=6cm,height=3.5cm]{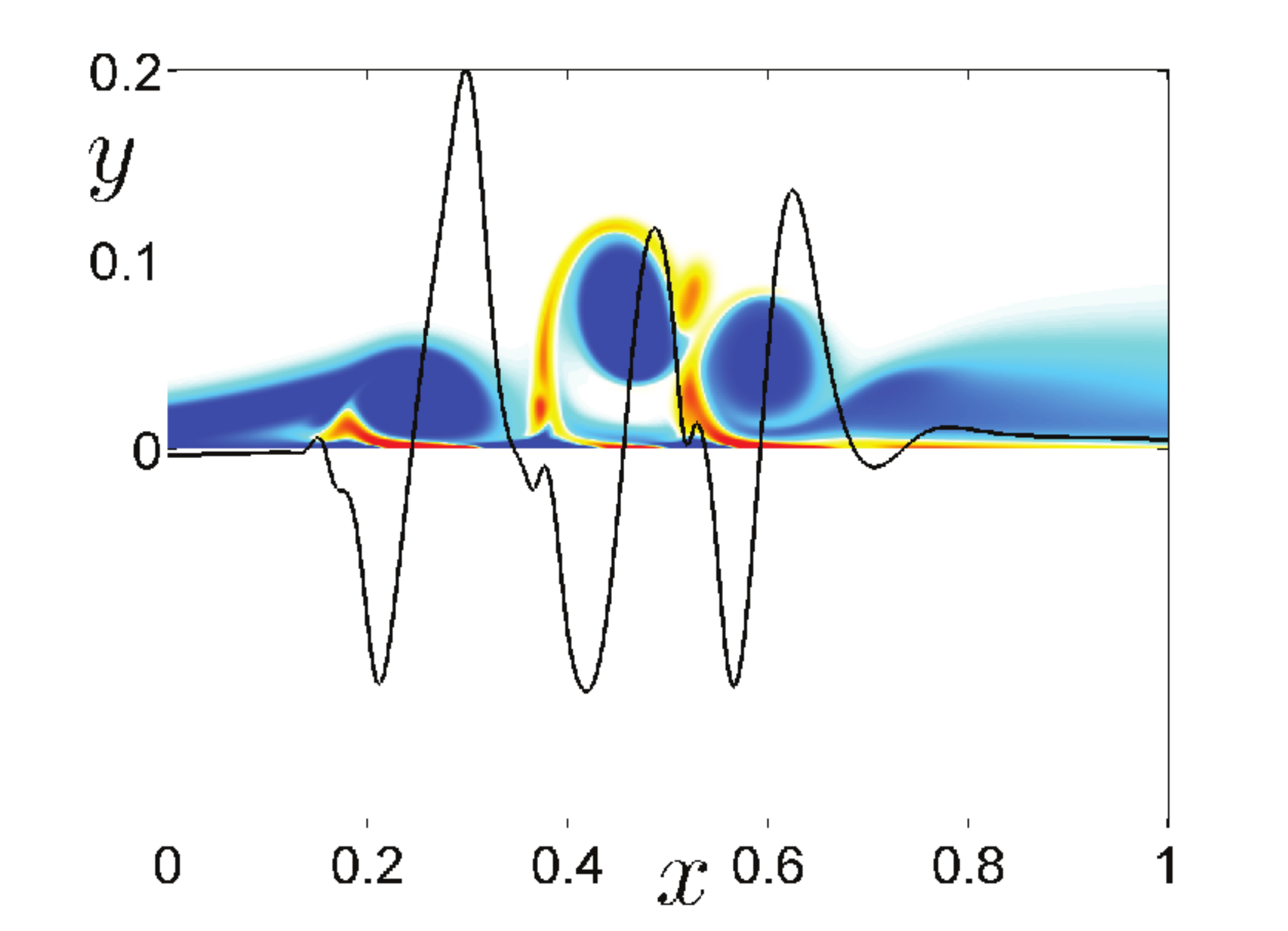}}
\subfigure[$Re=10^4,t=0.98$]{\includegraphics[width=6cm,height=3.5cm]{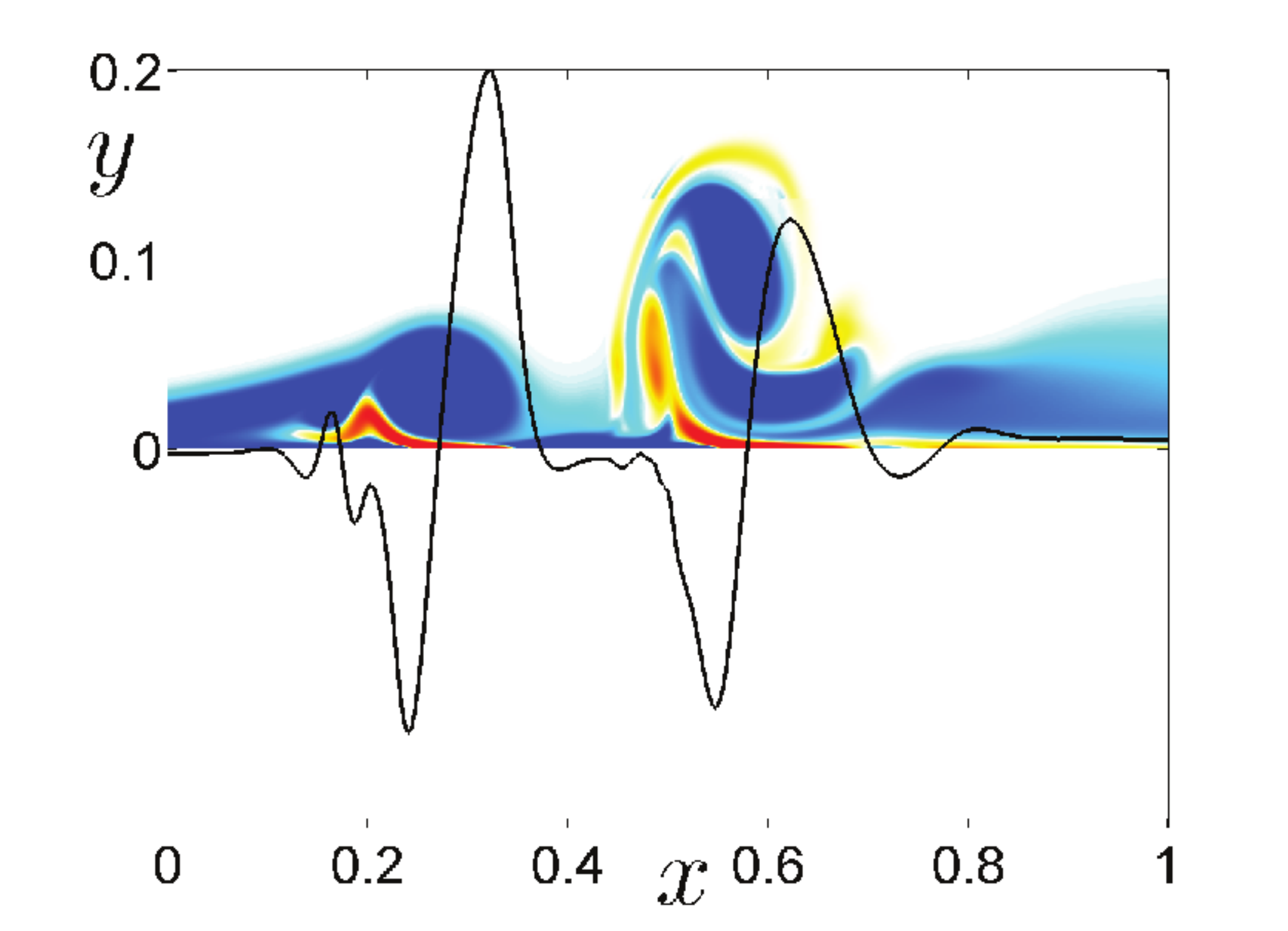}}
\caption{Vorticity contour levels for $Re=10^4$ at different times
compared with $\partial_x p_w$
(rescaled to fit the normal extension of vorticity). The blue colors represents
negative vorticity, red/yellow colors the positive vorticity.}
\label{vorticity10000}
\end{figure}
\begin{figure}
\centering
\subfigure[$Re=10^5,t=0.68$]
{\includegraphics[width=6cm,height=3.5cm]{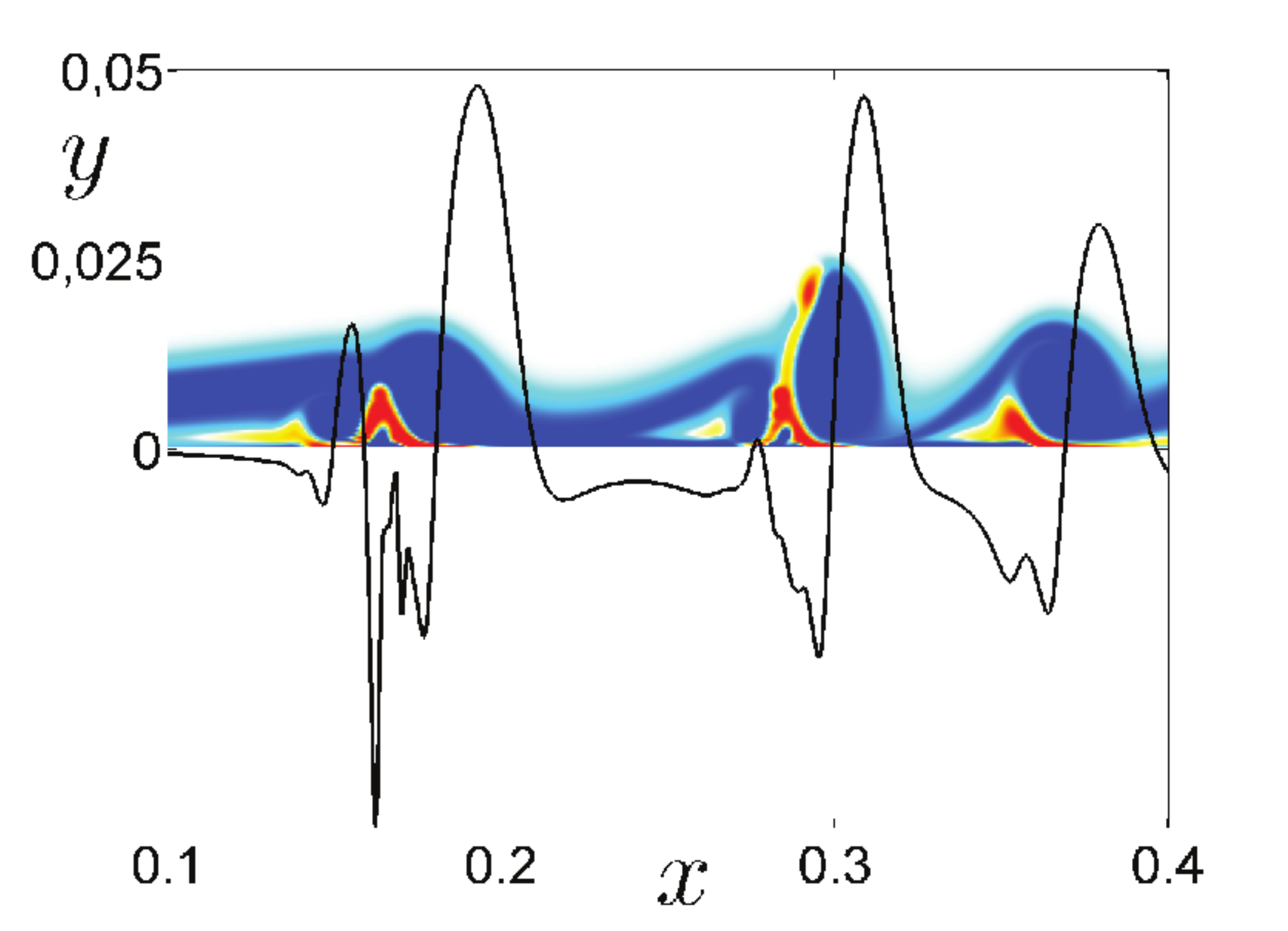}}
\subfigure[$Re=10^5,t=0.72$]
{\includegraphics[width=6cm,height=3.5cm]{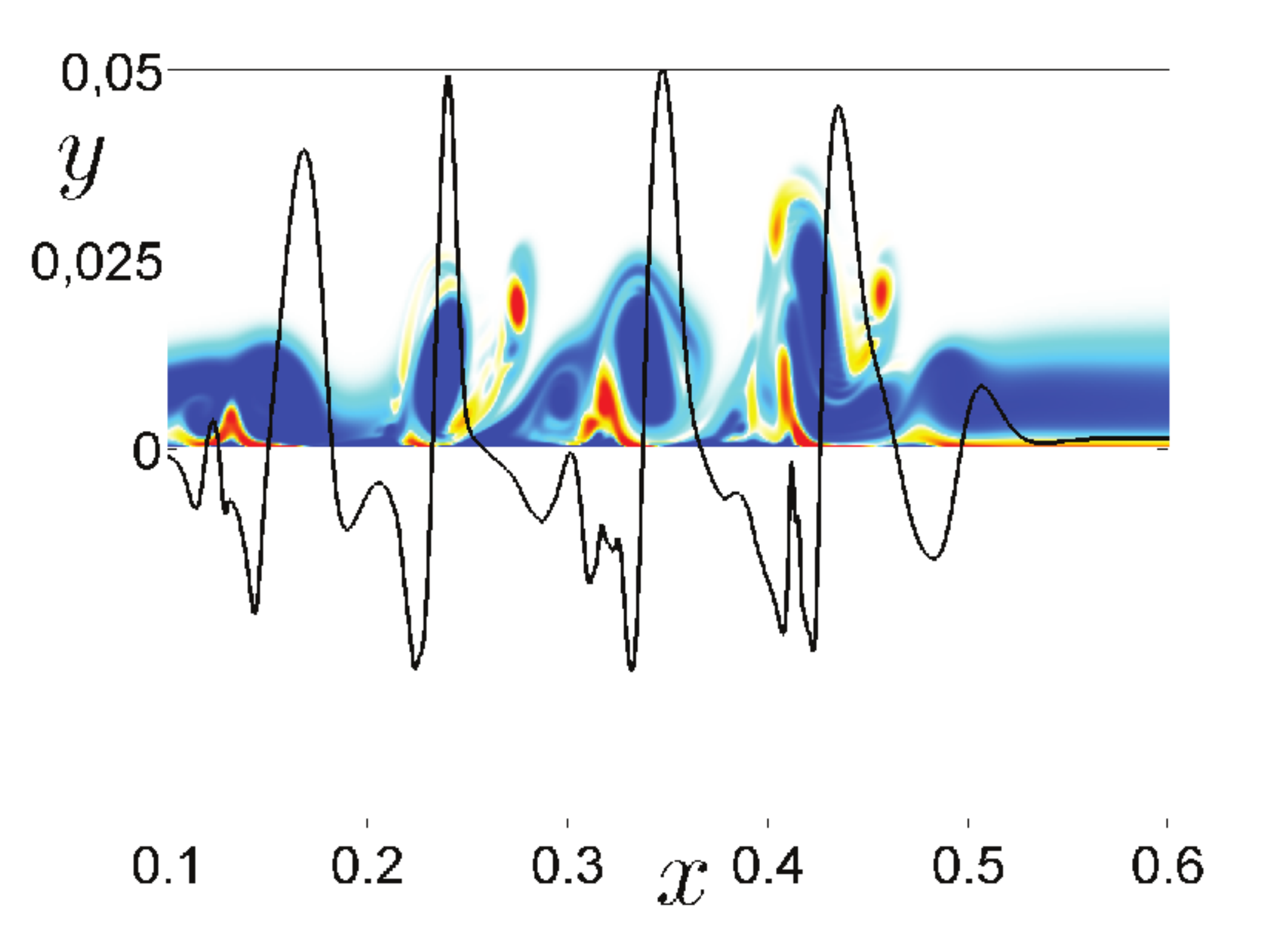}}
\caption {Vorticity contour levels for $Re=10^5$ at $t=0.68$ and $t=0.72$
compared with $\partial_x p_w$
(rescaled to fit the normal extension of vorticity). The blue colors represent
negative vorticity, red/yellow colors positive vorticity}
\label{vorticity100000}
\end{figure}

\subsection{Small-scale interaction: the formation of vortex dipoles}

The detachment process after the large-scale interaction strongly depends on the $Re$
number.
In Section 6.1  we have described the flow evolution for moderate-high
$Re$ numbers.
This regime is characterized by the evolution of the
large-scale interaction in the small-scale interaction and by the formation of
several recirculation regions on a small spatial scale.
To explain the physical features leading to small-scale interaction, we shall
focus mainly on the case $Re=10^4$, as it is simpler to describe than
$Re=10^5$.

\begin{figure}
\centering
\subfigure[Trajectory of vortex $B_{1-}$ for $Re=10^3,10^4$. The trajectory for
$Re=10^3$ is followed from $t=0.65$ to $t=1.5$ with temporal step of 0.2.
The trajectory for $Re=10^4$ is followed from $t=0.63$ to $t=1.05$ with temporal
step of 0.1.]{\includegraphics[width=10cm,height=5cm]{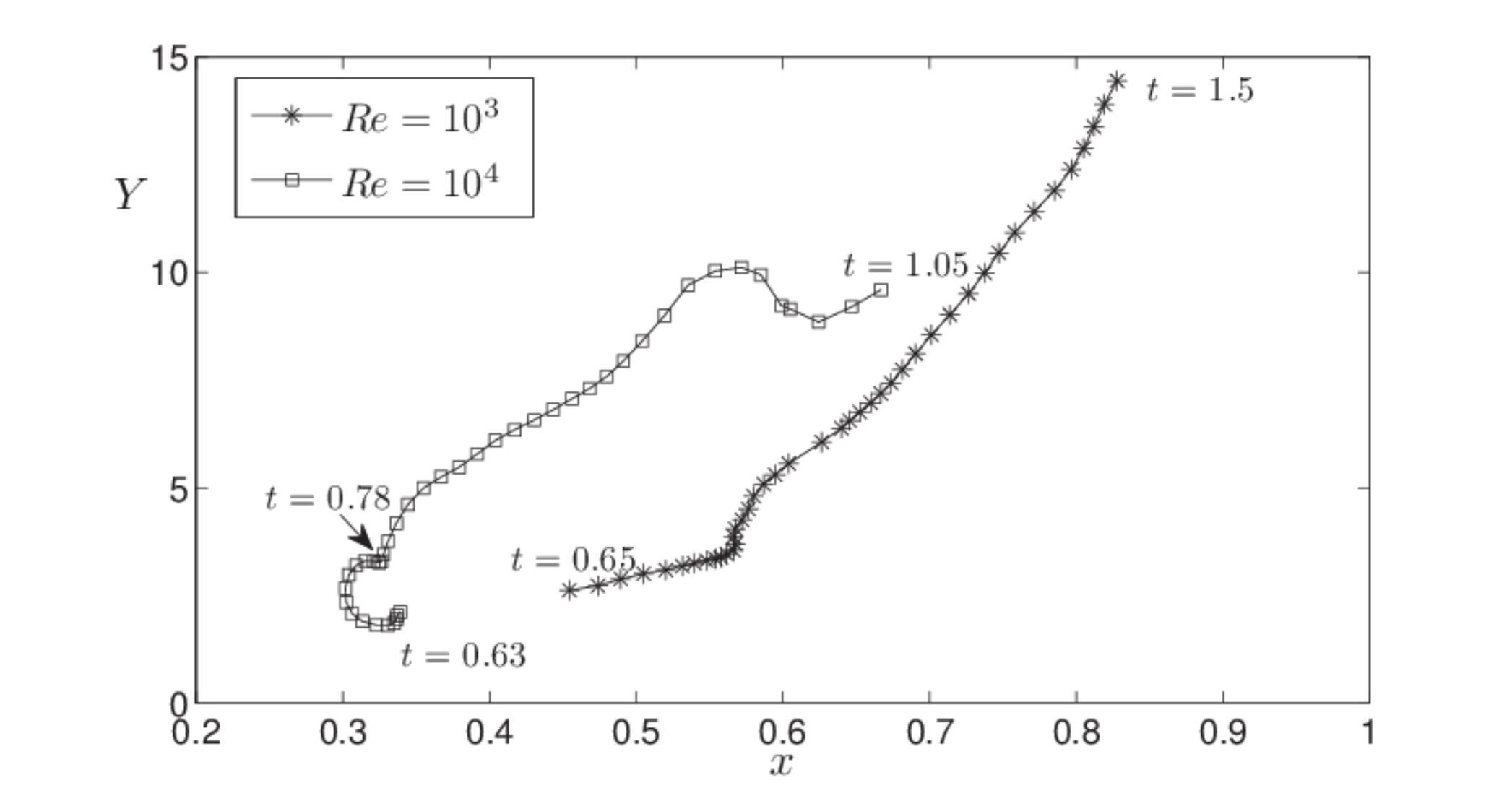}}
\subfigure[Trajectory of vortex $B_{2-}$ for $Re=10^3,10^4$. The trajectory for
$Re=10^3$ is followed from $t=0.95$ to $t=1.35$ with temporal step of 0.4.
The trajectory for $Re=10^4$ is followed from $t=0.7$ to $t=1.05$ with temporal
step of 0.1.]{\includegraphics[width=10cm,height=5cm]{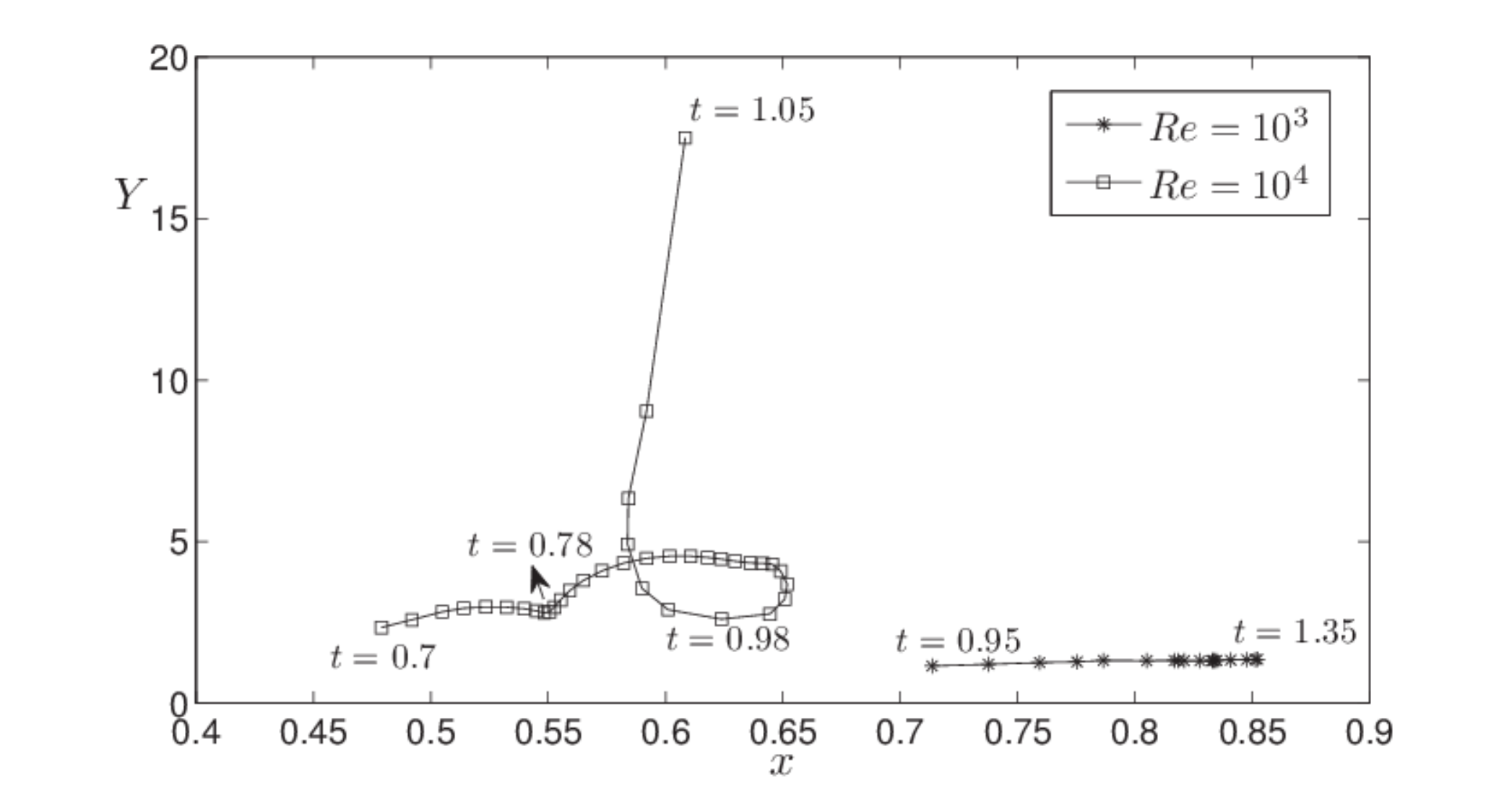}}
\caption{The trajectory of vortex $B_{1-}$ and $B_{2-}$ for $Re=10^3,10^4$. }
\label{B1m}
\end{figure}

\begin{figure}
\centering
\vspace*{-1.2cm}
\subfigure{\includegraphics[width=11cm,height=5.0cm]{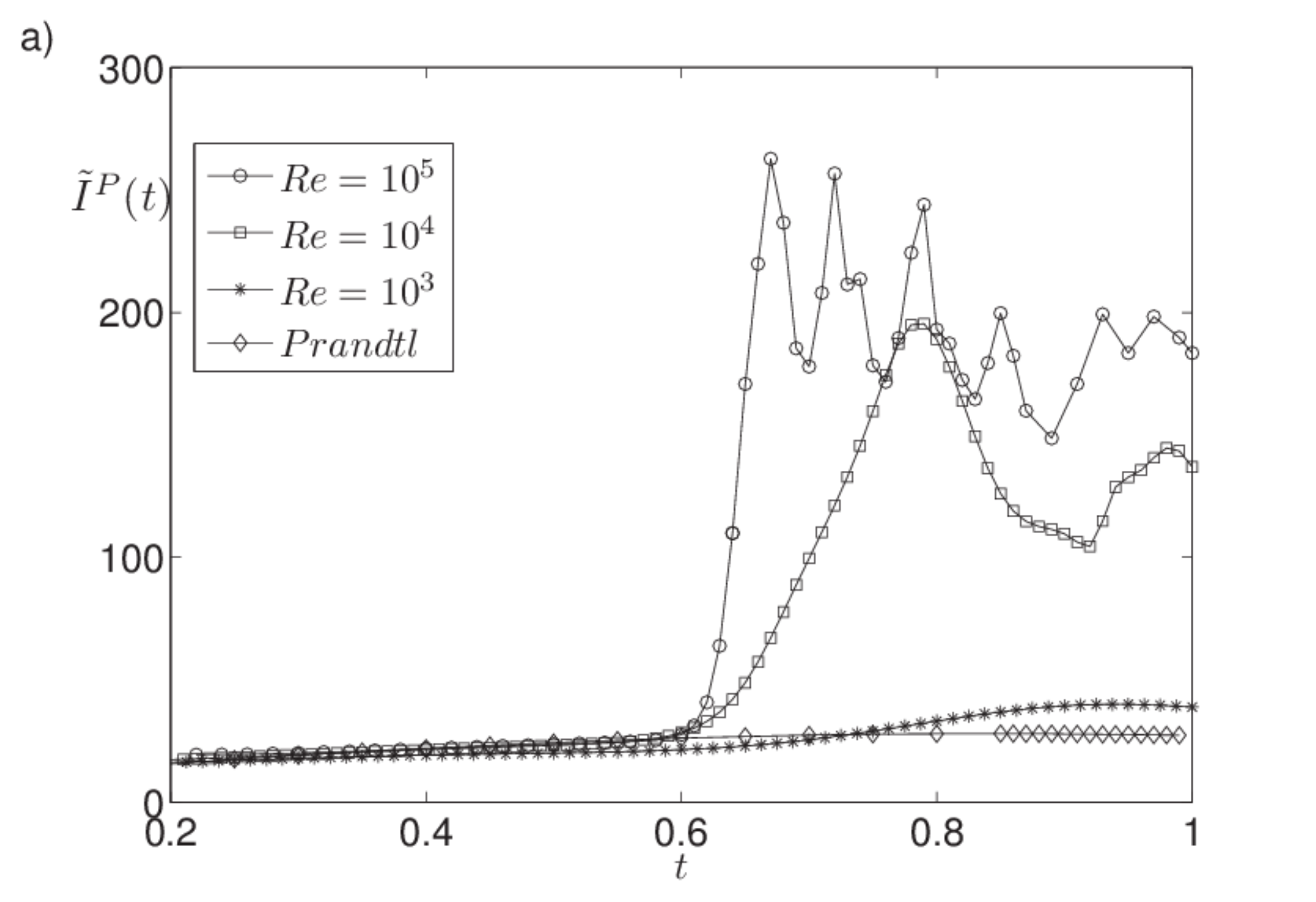}}
\vspace*{-0.4cm}
\subfigure{\includegraphics[width=11cm,height=5.0cm]{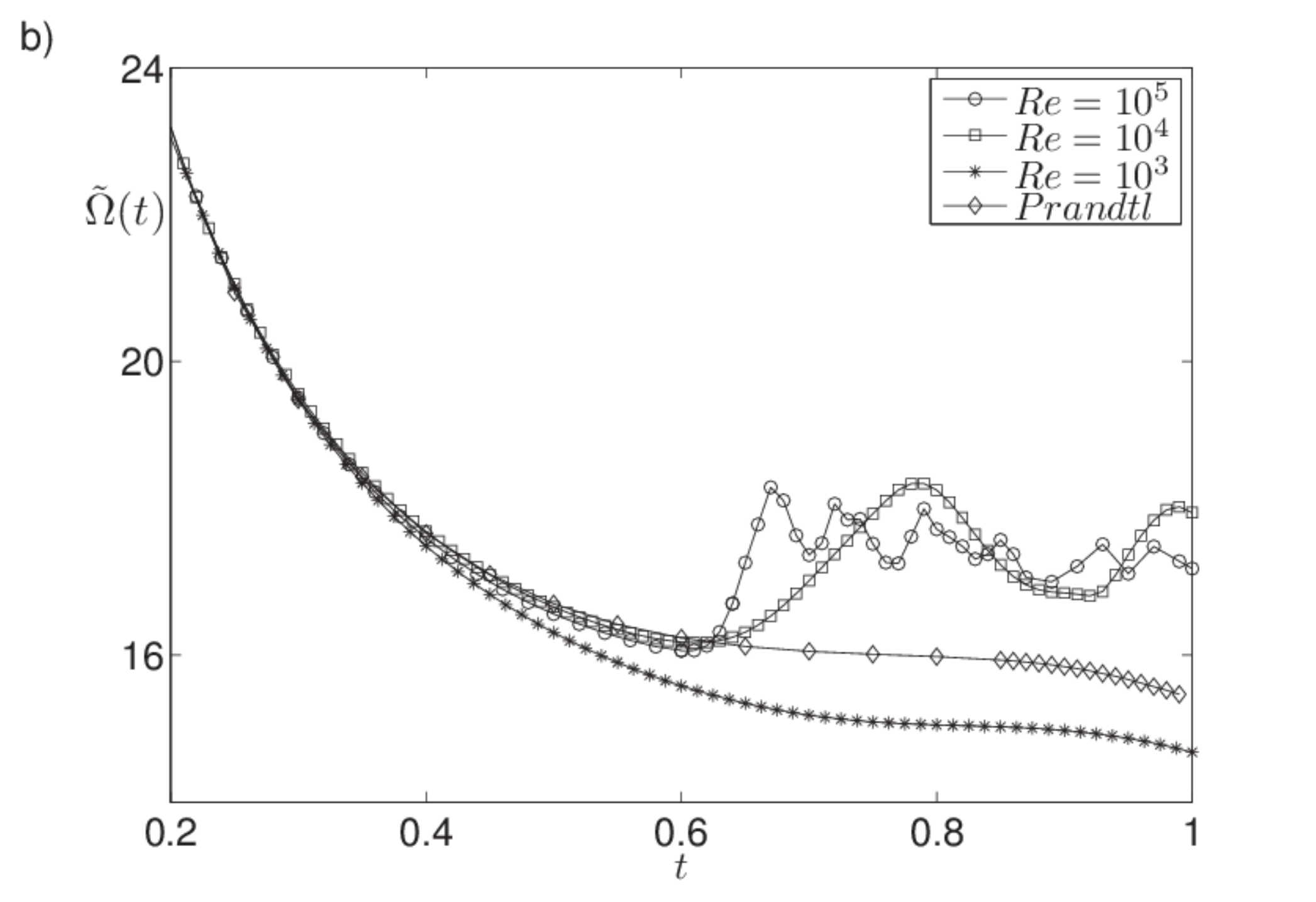}}
\vspace*{-0.4cm}
\subfigure{\includegraphics[width=11cm,height=5.0cm]{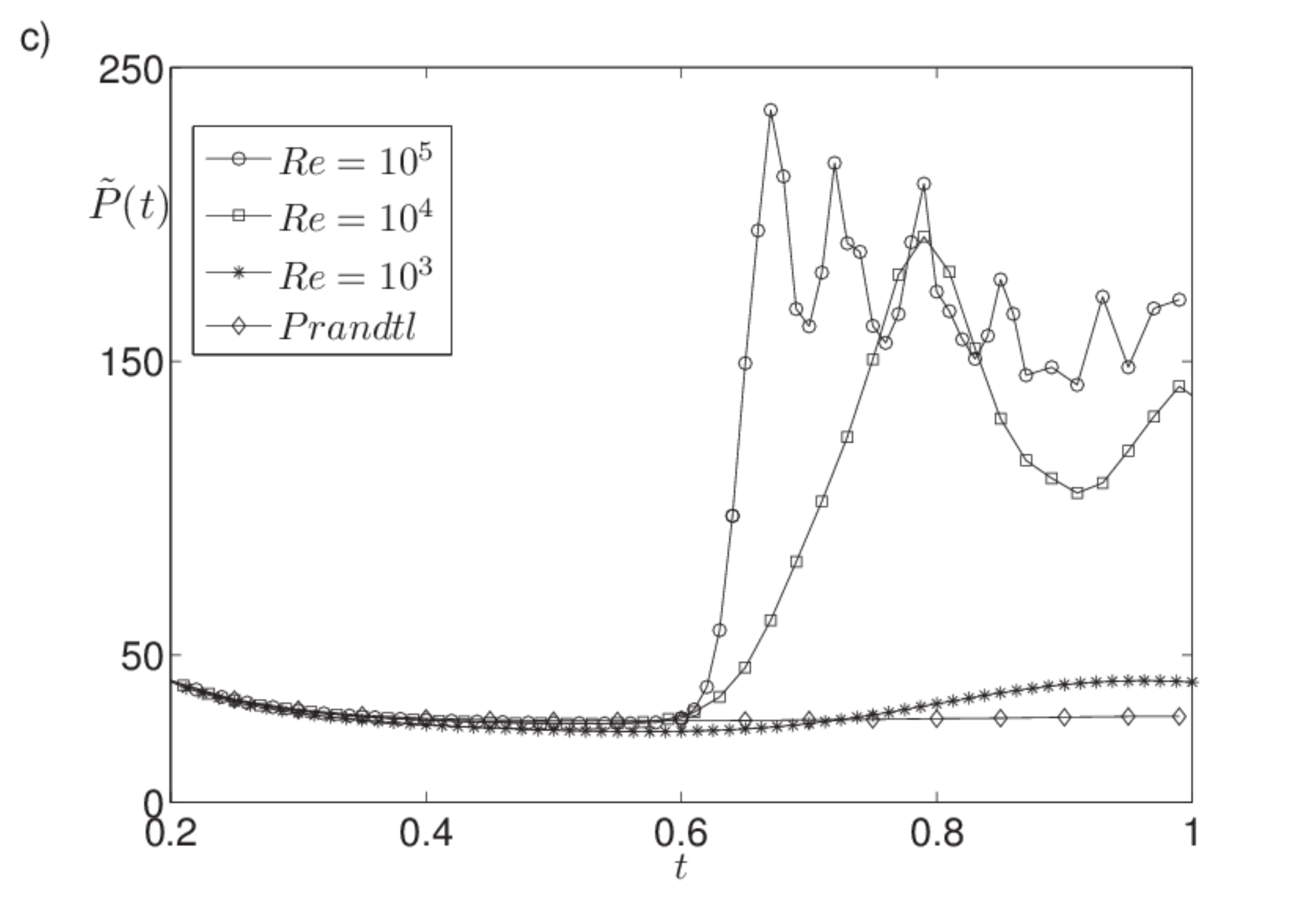}}
\vspace*{-0.4cm}
\subfigure{\includegraphics[width=11cm,height=5.0cm]{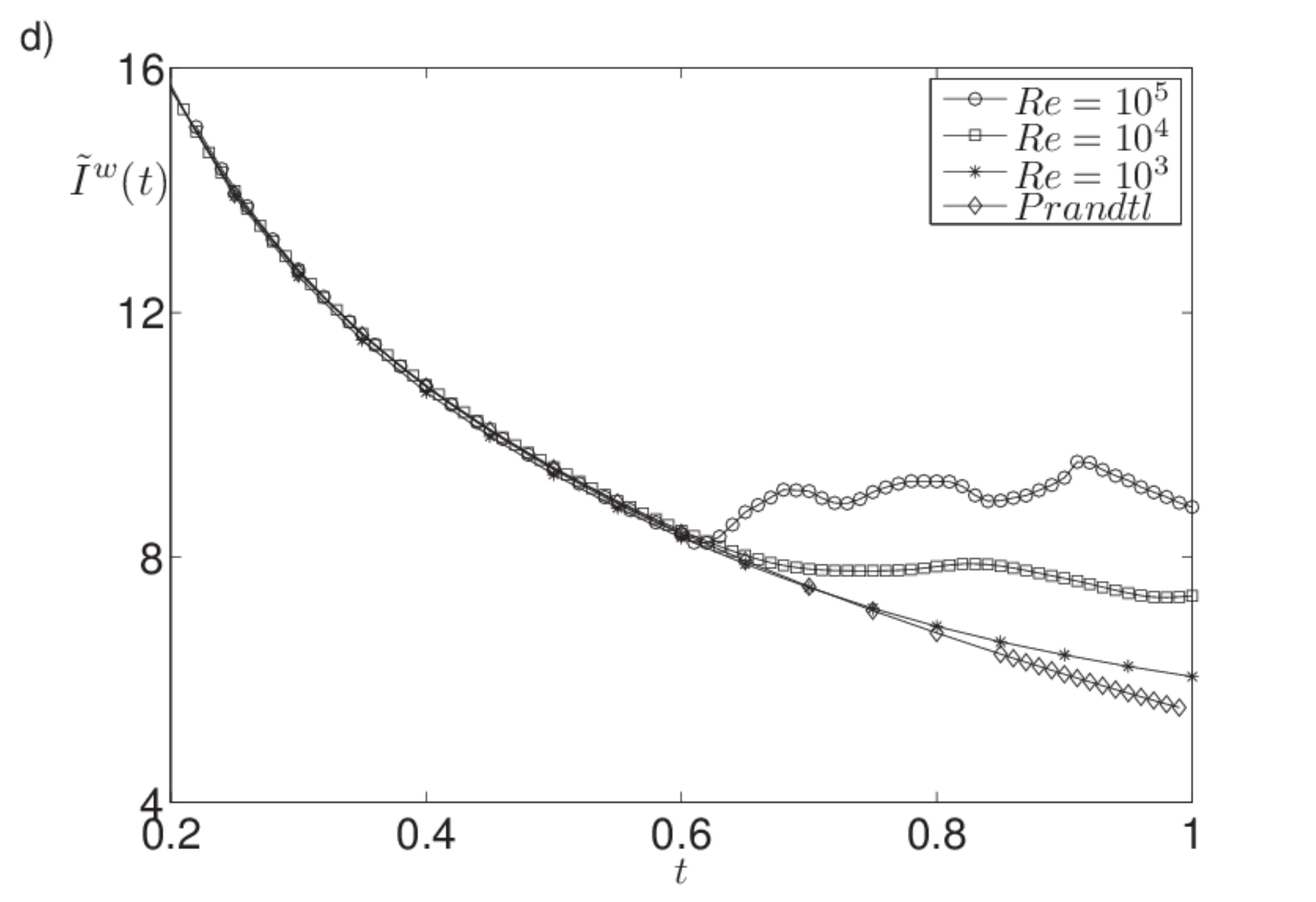}}
\vspace*{0.2cm}
\caption{A comparison between $I^p_P(t)$, $\Omega_P(t)$, $P_P(t)$, $I^w_P(t)$ and $I^p(t)$, $\Omega(t)$, $P(t)$, $I^w(t)$ at different $Re$ rescaled according to \eqref{rescaling}. Up to large-scale
interaction the good comparison reflects the good agreement between NS and Prandtl's solutions.
During the small-scale interaction (only for $Re=10^4-10^5$) the values for NS
strongly differ from Prandtl due to the interactions of the dipolar structures within the boundary layer.}
\label{enstrophyfig}
\end{figure}

From the detachment process of the boundary layer, a core of negative
vorticity $B_{1-}$ emerges and rotates clockwise
in the boundary layer. This core of negative vorticity forms at $t\approx0.63$
and it is clearly visible at $t=0.7$ in Fig.\ref{vorticity10000}a  centered in
$(0.29,0.023)$.
The vortex $B_{1-}$ forms a dipolar structure with the positive vorticity
$b_{1+}$, creating a favorable condition for the vorticity
production at the wall.
It is interesting to notice how $B_{1-}$ has enough strength to give a spin to
the positive vorticity $b_{1+}$,
which elongates around  $B_{1-}$ and penetrates in the zone of negative vorticity,
see Fig.\ref{vorticity10000}b.
During this movement, $b_{1+}$ breaks in two parts; the weaker part, that
from now on we shall denote with $b_{2+}$, remains connected to the wall.
At time $t\approx0.68$, a new core of negative vorticity, $B_{2-}$
(visible in Fig.\ref{vorticity10000} for all time),
is created on the right of $B_{1-}$, and it
forms a second dipolar structure with $b_{2+}$.
Moreover, from the detachment process of the first dipolar structure, given by
the coupling of $B_{1-}$ and $b_{1+}$, a third dipolar structure
forms on the left of $B_{1-}$, at $t\approx0.75$, visible in
Figs.\ref{vorticity10000}b-d
at $t=0.78,0.9$ and $t=0.98$.
The mutual interaction of these
dipolar structures results in the movement toward the wall of the negative
vortex cores $B_{1-}$ and $B_{2-}$.
This movement has a striking effect on the evolution in time of $\Omega(t)$ which
rapidly increases showing a first peak at $t\approx0.78$,
see Fig.\ref{enstrophyfig}b.
In fact at this time  the distances from the wall of the centers of $B_{1-}$ and $B_{2-}$ (see
Figs.\ref{B1m}a-b) reach a local minimum, and the positive
vorticity $b_{1+}$ and $b_{2+}$ are squeezed under $B_{1-}$ and $B_{2-}$,
leading to the production of a large amount of vorticity at the wall with is
signaled by growth of $I^p(t)$ visible in Fig.\ref{enstrophyfig}a.

The next peak in $\Omega(t)$ form at $t\approx0.98$ when
the first dipolar
structure strongly interacts with the second and pushes it close to the wall, as
shown in Fig.\ref{vorticity10000}d near $x=0.6$.
At this time, the center of $B_{2-}$ (see Fig.\ref{B1m}b) reaches
a new minimal distance from the wall, squeezing the zone of positive
vorticity $b_{2+}$ under $B_{2-}$. Subsequently, the two dipolar structures
merge together and move upward to finally interact with the
primary vortex.

The phenomena characterizing the dynamic of flow evolution for $Re=10^4$ are
also visible for $Re=10^5$, but in this case the formation of vortices
and dipolar structures within the boundary layer is much more chaotic and
difficult to describe.
Also for $Re=10^5$ the peaks in $\Omega(t)$, shown in
Fig.\ref{enstrophyfig}b, correspond to the impingements of the dipolar structures
on the wall.
Similarly to the case $Re=10^4$, the first peak in $\Omega(t)$
corresponds to the time when $B_{1-}$ gets closer to the wall.
This event happens at $t\approx0.68$, when other several dipolar structures
are also present, as visible in Fig.\ref{vorticity100000}a. The second
peak of $\Omega(t)$ is reached  at
$t\approx0.72$, when, similarly to the case $Re=10^4$, a dipolar structure near
$x=0.45$ (see Fig.\ref{vorticity100000}b) is pushed close to the wall due to the
interaction with another dipolar structure; at this time
eight cores  of negative vorticity are clearly visible.
Hereafter $\Omega(t)$ reaches several other peaks, as a consequence
of the complicated dynamics of the flow.

The effects of the small-scale interaction are therefore  related to the
formation of the peaks in $\Omega(t)$. The first peak forms
earlier  for $Re=10^5$ than for $Re=10^4$, confirming that the
large-scale interaction accelerates the small-scale interaction formation as the
$Re$ number increases.

The flow evolution for the case $Re=10^3$ is totally different from the
cases $Re=10^4,10^5$. In fact, for $Re=10^3$, the  viscosity  makes the
primary dipolar structures  weaker than the ones observed in the cases $Re=10^4,10^5$.
Like in $Re=10^4,10^5$, a core $B_{1-}$ of negative vorticity emerges within the
boundary layer at $t\approx0.65$, forming the first dipolar structure with the
positive vorticity zone $b_{1+}$. This structure is clearly visible in
Fig.\ref{vorticity1000}.
However, because of its weakness, $B_{1-}$ is not able to elongate
the positive vorticity $b_{1+}$ around $B_{1-}$, as seen for
$Re=10^4-10^5$, and this also prevents the break up of $b_{1+}$ and the subsequent formation of
other dipolar structures.
In fact, even if two other cores of negative vorticity $B_{2-}$ and $B_{3-}$ emerge
on the right and on the left of $B_{1-}$ at $t\approx0.95$ and $t\approx1.06$ respectively, they cannot
pair with other cores of positive vorticity,
preventing the interaction between the dipolar structures that
were responsible for the formation of the peaks in $\Omega(t)$ for
$Re=10^4-10^5$.

In Fig.\ref{B1m} one can also observe how the centers of $B_{1-}$ and $B_{2-}$
never move toward the wall, preventing the growth of $\Omega(t)$.
At time $t=1.5$, $B_{1-}$ has significantly moved away from the wall and it is going to
interact with the primary vortex.
One can  observe that the only relevant effects related to the formation of $B_{2-}$
are the small peaks in $I^p(t)$ and $P(t)$  that, however, combined in
\eqref{enstrophy} are not sufficient to increase the
enstrophy in the boundary layer.

In Fig.\ref{enstrophyfig} we report the enstrophy and the palinstrophy (and the related quantities
$I^p$ and $I^\omega$) of NS solutions rescaled as:
\be
\tilde{I}^p= I^p/Re^{1/2} ,\;  \tilde{\Omega}= \Omega/Re^{1/2} ,\;
\tilde{P}= P/Re^{3/2}, \tilde{I}^\omega= I^\omega Re^{1/2}  \, . \label{rescaling}
\ee
as well the same quantities for Prandtl's solutions.

The physical condition leading to the growth in time of the enstrophy is
therefore the impingement on the wall of the negative part of the dipolar
structures, similarly  to what has been shown in \cite{CB06,CH02,KCH07}
and previously in \cite{CL91,OR90},
where the authors studied the interaction of a vortex dipole with a
no-slip boundary.
This set-up differs substantially from the case described in this paper
and from the  case of the thick core vortex analyzed in \cite{OC02}.
In fact, if the dipole is sufficiently far from the wall, the flow
evolution does not differs noticeably from the free-slip or the stress-free case
and a very weak boundary-layer forms at $t=0$.
The boundary layer detachment derives from the movement toward the wall of the
dipole, which also leads to the formation of an adverse pressure gradient at the
wall, differently from our case where the adverse pressure gradient is instantly
imposed by the primary vortex.
However this detachment process also shows significant similarities with our case.
In fact in \cite{KCH07}, two different $Re$ number regimes were detected:
for $Re>O(10^4)$  a shear instability forms in the boundary-layer before it detaches from the wall,
leading to the roll-up of the boundary layer and to the formation of small-scale
vortex-structures, with large amount of vorticity
production.
For $Re<O(10^4)$, instead, no small-scale interaction was detected, and the
boundary layer detaches from the wall forming single vortices
which totally wrap-around the dipole halves.
Moreover for all the $Re$ regimes, during the
various rebounding of the dipolar structures on the wall,
the enstrophy increased when the dipoles get close to the wall,
similarly to what happens in our case.

%\\\\\\\\\\\\\\\\\\\\\\\\\\\\\\\\\\\\\\\\\\\\\\\\\\\\\\\\\\\\\\\\\\\\\\\\\\\\\\
\section{Conclusions}
We have computed the solutions of 2D Prandtl and Navier-Stokes equations
in the case of a rectilinear vortex interacting with a wall.
We have analyzed the asymptotic validity of boundary layer theory  by comparing
Prandtl's solution with the NS solutions for $Re$ in the range $10^3-10^5$.
In our case Prandtl solution  terminates  in a singularity at time $t\approx0.989$.
The singularity formation is anticipated by a first interaction of the boundary
layer flow with the outer flow, which is revealed by the spiky behavior
of the streamlines and of the displacement thickness at $t\approx0.85$.
This is the consequence of the compression of the flow in the streamwise direction
in a very narrow zone, which leads to an eruption in the normal direction with
ejection of flow from within the boundary layer to the outer flow at singularity time.

The unsteady separation process, as predicted by the NS equations, has a different
evolution, at least for the  $Re$ regimes considered here.
In fact, we have seen  a good quantitative agreement between Prandtl and NS solution
only during the earlier stages, until the beginning of the large-scale
interaction, which is signaled by the local change of the streamwise pressure gradient
in the boundary layer.
This interaction starts later the higher the  $Re$  number,  and acts over the flow
in a region close to the boundary whose size, in the streamwise direction, is
comparable with the size of the recirculation region.
We set as the beginning of large-scale interaction the formation of an inflection
point in the streamwise pressure gradients on the wall, as the relative change
of concavity represents a different topological structure as compared to the
streamwise pressure gradients imposed in classical BLT.

The character of the large-scale interaction is different from the interaction that
arises in boundary layer theory before the singularity time, as no large gradients
nor spike-like structure is visible  in the Navier-Stokes solution during that stage.
However this interaction can be considered the precursor of the following
stage of the evolution which is characterized by the small-scale interaction.

The small-scale interaction (which we have seen for $Re=10^4,\, 10^5$, while it is
absent for $Re=10^3$) is revealed by the formation of a spike in the
solution of NS equations, and in this sense it is reminiscent of the singularity
developed in Prandtl's solution.
This stage of the separation process
is characterized by the formation, within the boundary layer, of several recirculation
regions and  dipolar vortical structures, leading to a complicated
flow dynamics revealed also by the growth of the enstrophy.
In our simulations the small-scale interaction occurs  prior to the time
of spike formation in classical BLT.
However we note that in the NS solutions the small-scale interaction
stage manifests slightly earlier
as the $Re$ increases, supporting the conjecture presented in \cite{Cas00,OC02}
for the case of the unsteady separation induced by a thick-core vortex,
according to which the large-scale interaction accelerates the small-scale
interaction formation as $Re$ increases.
Another feature of the small-scale interaction is the growth of the normal pressure
gradient inside the BL; this growth causes the normal pressure gradient to become
an $O(1)$ quantity, and reveals another important departure of the NS solution
from the classical BLT.

A striking (and very simple to be revealed) effect of the small sale phenomenology
is the growth in time of the enstrophy of the flow inside the boundary layer.
This growth, caused by the collision on the wall of the dipolar structures
that forms during the separation process, is absent both in Prandtl solutions
as well in NS solutions for low $Re$, and has noteworthy similarities with the phenomena
analyzed  in \cite{CB06,CH02,CL91,KCH07,OR90} for the case of the interaction
of a dipole with a wall.

An important point would be to investigate the behavior for higher $Re$ number regimes.
The  results presented  in the present paper (as well the results of \cite{OC02})
might suggest the possibility that, for very high $Re$ no large-scale interaction
occurs and the interaction between the boundary layer with the outer flow manifests
only in terms of small-scale interaction.
Given the lack of the large scale interaction (which, as noted before, accelerates
the beginning of the small-scale stage) this would delay the beginning of the
small-scale interaction which would become closer to the Prandtl's
singularity time the higher the $Re$ is.
Another scenario would be that Rayleigh-type instabilities manifest at a time
prior to the Prandtl's singularity time; the possibility that instability  wins
the race with Prandtl's singularity was raised in \cite{Cow01} and seems supported
by the computations of \cite{OC10}.
For $Re=10^6$ we have detected the same phenomenon, but at this stage and for the
resolutions we have been able to attain, it is difficult to discern between spurious
numerical instability and physical Rayleigh instability.
This topic will be the subject of future work.

\bibliographystyle{siam}
\bibliography{prandtl}

\end{document}